\documentclass{aastex6}

\usepackage[hyperref]{} 
\usepackage{epsfig}
\usepackage{psfig}
\usepackage{graphicx}
\usepackage{epstopdf}

\renewcommand{\aap}{    {\it Astron. Astrophys.}}

\renewcommand{\apj}{    {\it Astrophys. J.}}
\renewcommand{\apjl}{   {\it Astrophys. J. Lett.}}

\renewcommand{\jgr}{    {\it J. Geophys. Res.}}

\renewcommand{\solphys}{{\it Solar Phys.}}

\newcommand{\be}{\begin{equation}}
\newcommand{\ee}{\end{equation}}
\newcommand{\bea}{\begin{eqnarray}}
\newcommand{\eea}{\end{eqnarray}}
\newcommand{\beax}{\begin{eqnarray*}}
\newcommand{\eeax}{\end{eqnarray*}}
\newcommand{\ba}{\begin{array}}
\newcommand{\ea}{\end{array}}
\newcommand{\bed}{\begin{description}}
\newcommand{\ed}{\end{description}}
\newcommand{\blc}{\begin{list}{$\circ$}{}}
\newcommand{\blb}{\begin{list}{$\bullet$}{}}
\newcommand{\el}{\end{list}}
\newcommand{\ben}{\begin{enumerate}}
\newcommand{\een}{\end{enumerate}}

\def\gsim{\,\lower3pt\hbox{$\sim$}\llap{\raise2pt\hbox{$>$}}\,}
\def\lsim{\,\lower3pt\hbox{$\sim$}\llap{\raise2pt\hbox{$<$}}\,}

\begin{document}

\tracingmacros=2

\title{The Roles of Reconnected Flux and Overlying Fields in CME Speeds}

\author{M. Deng\altaffilmark{1}}
\affil{Department of Applied Physics, 348 Via Pueblo Mall, Stanford University, Stanford, CA 94305-4090}
\email{mindad@stanford.edu}
\and
\author{Brian T.~Welsch\altaffilmark{2}}
\affil{Natural \& Applied Sciences, University of Wisconsin - Green
  Bay, 2420 Nicolet Dr., Green Bay, WI 54311}
\email{welschb@uwgb.edu}

\altaffiltext{1}{Department of Physics, University of California,
366 LeConte Hall, MC 7300, Berkeley, CA 94720-7300} 
\altaffiltext{2}{Space Sciences Laboratory, 7 Gauss Way, University of
California, Berkeley, CA 94720-7450}





\shortauthors{M.Deng and B.T.Welsch}
\shorttitle{Reconnected Flux \& Overlying Fields {\it vs.} CME Speeds}



\begin{abstract}
Researchers have reported (i) correlations of coronal mass ejection
(CME) speeds and the total photospheric magnetic flux swept out by
flare ribbons in flare-associated eruptive events, and, separately,
(ii) correlations of CME speeds and more rapid decay, with height, of
magnetic fields in potential coronal models above eruption sites.
Here, we compare the roles of both ribbon fluxes and the decay rates
of overlying fields in a set of 16 eruptive events. We confirm
previous results that higher CME speeds are associated with both
larger ribbon fluxes and more rapidly decaying overlying fields.  We
find the association with ribbon fluxes to be weaker than a previous
report, but stronger than the dependence on the decay rate of
overlying fields.  Since the photospheric ribbon flux is thought to
approximate the amount of coronal magnetic flux reconnected during the
event, the correlation of speeds with ribbon fluxes suggests that
reconnection plays some role in accelerating CMEs. One possibility is
that reconnected fields that wrap around the rising ejection produce
an increased outward hoop force, thereby increasing CME acceleration.
The correlation of CME speeds with more rapidly decaying overlying
fields might be caused by greater downward magnetic tension in
stronger overlying fields, which could act as a source of drag on
rising ejections.
\end{abstract}

\keywords{Coronal mass ejections; Magnetic fields, corona; Magnetic fields, models}
\tracingmacros=0

\section{Introduction}
\label{sec:intro}

In a coronal mass ejection (CME), magnetic forces in the low corona
accelerate a hot ($\approx 1$ MK) mass ($\approx 10^{15}$ g) of magnetized
plasma at high speed (from a few hundred km sec$^{-1}$ at the slow end
to 2000 km sec$^{-1}$ or more at the fast end) into interplanetary
space. These events are the primary drivers of severe space weather
disturbances at Earth \citep{Gosling1993}, but key aspects of their
initiation and subsequent evolution are not well understood.
Prediction of CME speeds prior to their arrival at 1 AU is an
important goal for space weather forecasts, both to accurately
estimate their arrival times, and to account for the known correlation
between CME speed and geoeffectiveness \citep{Richardson2011}.

Characterizing the forces that act upon CMEs as they begin to
propagate away from the Sun is essential to understand, and eventually
predict, their dynamics.  Here, we investigate two reported influences
on the dynamics of flare-associated CMEs: (i) magnetic reconnection
during the eruption process, as quantified by the amount of
photospheric magnetic flux swept out by flare ribbons; and (ii)
magnetic tension from large-scale magnetic fields that overlie the
eruption site, parametrized by the rate of decay of horizontal field
strengths with height, $r$, above the photosphere in potential field
models.

In the ``standard model'' of eruptive flares, dubbed the CSHKP
(Carmichael-Sturrock-Hirayama-Kopp-Pneumann; {\it e.g.}, \citealt{Webb2012,
  Svestka1992}) model, 
%
%
magnetic reconnection in the corona beneath the rising ejection leads,
either directly or indirectly, to particle acceleration.  The
resulting energetic flare particles then propagate downward along the
newly-reconnected magnetic fields toward the transition region and
upper chromosphere, where they interact with the denser plasma and
  heat the atmosphere.  
This heating then powers dramatic intensity enhancements, which
typically form two parallel, elongated emission structures, known as
flare ribbons.  These ribbons always straddle polarity inversion lines
(PILs), along which the radial photospheric magnetic field changes
sign, which is consistent with ribbon emission occurring on conjugate
footpoints of newly reconnected magnetic flux.  In the simplistic
CSHKP picture, the erupting structure itself originates directly above
the PIL.  Flare ribbons tend to systematically move apart as the
  ejection rises, as the reconnection links fields with footpoints
  that are farther and farther apart.  Observers often study ribbons
in H$\alpha$ and UV images ({\it e.g.}, in the 1600 \AA\, channel of the
Transition Region and Coronal Explorer (TRACE) satellite \citep{Handy1999}.

Flare ribbons overlie strongly magnetized areas of the photosphere
within and near active regions (ARs), and it was long ago hypothesized
that the rate of (unsigned) magnetic flux being swept out by the
ribbons as they move apart should be related to the rate of coronal
magnetic reconnection \citep{Forbes1984, Poletto1986}. More recently,
in a study of 13 CMEs with flares, \citet{Qiu2005} reported a very
strong correlation between the total unsigned magnetic flux swept out
by ribbons over the course of each event --- which we hereafter refer
to as the ribbon flux --- and the CME speed: they found a linear
correlation coefficient of 0.89. In a related study, \citet{Qiu2004}
reported an association between the {\it rate} of flux being swept out
by ribbons and CME acceleration.  These findings are consistent: the
time integral of acceleration yields velocity, and the time integral
of the rate of magnetic flux being swept by ribbons ($\propto$
acceleration) yields the total magnetic flux swept by ribbons
($\propto$ velocity).

The large-scale magnetic environment of CME source regions is also
believed to influence CME dynamics.  
\citet{Kliem2006} argued that CMEs are a manifestation of the torus
instability, by which a toroidal current ring will be unstable to
expansion if the external field surrounding the torus --- which could inhibit
any expansion via magnetic tension --- decreases sufficiently rapidly
in strength with radial distance from the center of the torus.
Parametrizing the rate of decrease in external field strength with
radius by defining a decay index, they theoretically characterized the
dependence of the acceleration of the outward-expanding torus on the
decay index, finding that a more rapid decay of external fields
implies a more rapid expansion.
\citet{Demoulin2010} analytically modeled instabilities in more
general configurations in terms of decay indices, and found that
instability thresholds depended upon the field's structure.
Observationally, \citet{Wang2007b} and \citet{Liu2008} suggested that
magnetic fields lying above flare sites can inhibit eruptions: using
potential-field, source-surface (PFSS) models of magnetic fields above
flare sites, they found that flares below horizontal fields that
remained relatively strong with increasing height were less likely to
be eruptive than flares below horizontal fields that decayed more
rapidly with height.  This is consistent with the idea that magnetic
tension from stronger fields above flare sites is more capable of
confining would-be ejections than weaker fields.  \citet{Liu2008}
quantified the rate of decay of magnetic field with height using a
decay index defined to be the fitted index of an assumed power-law
decrease in horizontal magnetic field strength with radial height $r$
above the flare site in the interval between about 45 Mm and 115
Mm above the Sun's surface.

Beyond the role of the rate of decay of overlying magnetic fields in
the genesis of CMEs, overlying fields might also influence the speed
of a CME after it has begun.
When eruption occurs, it is physically plausible that magnetic tension
from magnetic flux overlying a CME source region that is entrained by
the leading edge of a rising ejection could slow the resulting CME.
Also, flux above an eruption site that is not entrained would still
need to be pushed aside, and might thereby impose an ``effective
drag'' on the erupting CME, since this pushing would require doing work
against magnetic pressure.
Consistent with this picture, \citet{Xu2012} studied 38 CMEs and
compared plane-of-sky CME speeds from the {\it Large Angle
  Spectroscopic Coronagraph} (LASCO; \citealt{Brueckner1995}) CME
catalog with decay indices derived from local, Green's function
potential field extrapolations above manually identified PILs.  They
reported that faster CMEs had larger indices; that is, field strengths
above faster CMEs' source regions decreased more rapidly with
height. Although they did not compute correlation coefficients between
their decay indices and CME speeds, their table contains the necessary
information to do so, and we find linear and rank-order correlations
of 0.64 and 0.60, respectively.  We note that their extrapolations
used evenly spaced grid points in height, but they apparently fitted
the logarithm of height, so their fitted decay indices are probably
biased toward the rate of change in field strengths at larger heights.
To the extent that fields decay more rapidly at greater heights, as in
their Figure 1, this might bias their decay indices to be
systematically larger than fits with points evenly spaced in the
logarithm of height.

We note that the approach of only modeling the field directly above
the flaring PIL followed by \citet{Wang2007b}, \citet{Liu2008}, and
\citet{Xu2012} does not account for the effects of complex magnetic
structure in the erupting field.  For instance,
\citet{Sterling2001c} consider a scenario in which CMEs erupt sideways,
from the edge of an active region.
We also remark that the potential-field models that we analyze lack
any electric currents, so must certainly differ from the actual
coronal fields, which evidently possess electric currents that store
the energy needed to power CMEs.

Based upon these reports of the roles of ribbon fluxes and decay
indices in CME speeds, we set out to investigate the relative
strengths of these two factors in determining CME speeds.  Which
factor matters most?  And what is the unexplained variance in CME
speeds after accounting for the joint influence of both factors?

To address these and other questions, we analyze both ribbon fluxes
and decay indices in a sample of 18 CMEs.  In the next section, we
present the events in our sample.  Then, in Section \ref{sec:methods},
we describe our methods for analyzing each event, such as identifying
ribbons and PILs, extrapolating fields, and determining decay indices.
In Section \ref{subsec:case}, we illustrate these in a case study.  We
then present our results in Section \ref{sec:results}, and conclude
with a discussion their significance in Section \ref{sec:discussion}.


\section{Data} 
\label{sec:data}

Once again, our aim is to quantitatively characterize the joint
influence of ribbon flux and decay rate of overlying fields on CME
speeds.  To do so, we study a sample of 16 CME events from a seven year
span, from 1998 to 2005.  In Table \ref{tab:data}, we show event
times, solar source locations, CME speeds, and other parameters
discussed below.

In order to be able to compare our results with the findings of
\citet{Qiu2005}, we analyze several events from that study. All the
eruptive events we chose were ``halo'' CMEs, meaning that all CMEs were
Earth-directed. Usually, halo CMEs originate from source active
regions near disk center, so from the perspective of Earth-based
observations, we can study them in greater
detail \citep{Webb2012}. More significantly, since halo CMEs are
responsible for the strongest geomagnetic storms and other space
weather phenomena \citep{Webb2012}, it is very important for us to be
able to predict them. Also, the sample we chose consists of mostly
fast CMEs: only four have speeds less than 800 km s$^{-1}$. We note
that there are two events from the same source region, AR 10486,
separated by several hours. These events are very famous, with
remarkably high CME speeds ($>$ 2000 km s$^{-1}$) and ribbon fluxes.

Ribbon fluxes and magnetic field maps in event source regions for all
of these events were provided by Qiu and collaborators
(\citealt{Qiu2005, Qiu2007}, and private communication). Full-disk,
line-of-sight (LOS) magnetograms for these events were recorded by the
{\it Michelson Doppler Imager} (MDI; \citealt{Scherrer1995}), and Qiu
and collaborators extrapolated from the photospheric level to 2000 km,
to estimate the flux density near the expected height of formation of
ribbon emission.
Here, the LOS field was used directly as an approximate measure of the
radial magnetic field, and pixel areas were not corrected for
foreshortening from the viewing angle.
Images of flare ribbons were recorded either at Big Bear Solar
Observatory, in H$\alpha$, or by the TRACE satellite, in UV.
For each event, Qiu {\it et al.} generated 26 different ribbon masks ({\it i.e.,}
two-dimensional, binary maps) outlining the cumulative areas swept out by flare
ribbons over the course of each event, by slightly varying selection
criteria for ribbon pixels \citep{Qiu2005, Qiu2007}.
These variations were undertaken to enable estimation of systematic
errors due to mask selection parameters.
Magnetogram data were interpolated onto the grid corresponding to the
H$\alpha$/TRACE pixels.  (This introduced artifacts at the edges of
some magnetograms, which we discuss further below.)
The lack of adjustment for foreshortening introduces errors of $10 \pm
7$\% over the set of all event fluxes; variations resulting from
different ribbon-mask selection criteria are substantially larger.
Note that, following \citet{Qiu2005}, the ribbon flux is the total
unsigned magnetic flux swept by the ribbon pixels --- {\it i.e.}, by
including ribbon pixels in both polarities, the ribbon flux should be
approximately double the amount of reconnected magnetic flux.

\section{Models and Methods}
\label{sec:methods}

Since we already have estimates of the ribbon flux and CME speed for
each event, in order to compare the relative influence of decay
indices on CME speeds we must still estimate the how the magnetic
field changes with height above the CME source region for each event.

As an initial step, we developed an automated (and therefore
objective) method to define each CME's source region on the photosphere
from ribbon emission.  It should be noted that CMEs are large-scale
(possibly global) phenomena, so the small source region that we
identify for each event should be interpreted more as the center or
core of a much larger, surrounding source region than as an origin of
all CME-associated dynamics.  We expect that the magnetic field
structure above the photospheric source region that we identify
represents the large-scale magnetic environment of the erupting CME.

As noted above, flare ribbons are thought to be the photospheric
manifestation of coronal reconnection, so indicate which PIL(s) within
the AR participated in the flare reconnection.  Therefore, much as
\citet{Xu2012} did, we used ribbon loci to determine, in part, where
to model the structure of coronal fields.  While ribbons typically
straddle the PIL, the eruption itself is assumed to originate above
the PIL; hence, we define CME source areas to be near-PIL regions that
are bracketed by flare ribbon emission.
%
To identify near-PIL regions, we used an algorithm similar to that of
\citet{Welsch2008b}, which requires both positive and negative fields
with unsigned flux densities above a threshold of 100 Mx cm$^{-2}$ to
be close together: single-polarity magnetograms were created, then
dilated by five pixels, and multiplied together; nonzero areas in the
result correspond to the PIL mask.
This flux-density threshold limits our analysis to strong-field
PILs, by excluding insignificant PILs from small-scale fluctuations in
weak-field regions.  We then found the intersection of these near-PIL
regions with masks of ribbon emission.
In Section \ref{subsec:case}, we study how varying parameters used in
defining near-PIL regions can effect our estimates of the rate of
decay of overlying magnetic fields.

In addition, for some of the events, (i) strong-field PILs and (ii)
transient emission enhancements both occur far from the main flaring
PIL(s).  While this spatially remote, transient emission is likely
related to the flare, we do not expect the overlying magnetic fields
above such distant regions to strongly influence dynamics of the erupting
structure near the core ribbon areas.
Here, we have implicitly assumed that the bulk of the CME originates
from above the main flaring PIL.  We justify this assumption via
Occam's razor: flare ribbons are known to be associated with
eruptions; we have no evidence that the eruptions that we study
originated from parts of their source ARs remote from the largest area
of ribbon emission; we therefore favor the simplest assumption given
the observations, {\it i.e.}, that the CMEs that we study originate in the
corona above the main site of ribbon emission.  Consequently, we
choose to exlude fields overlying far-away, small-scale brightenings
%
%
from our analysis.  To do so, we kept only those pixels within a
proximity threshold of 75 pixels from the geometric centroid of the
ribbon mask for each event.
Note that because pixel sizes vary between the interpolated
magnetograms used in our extrapolations, this threshold does not
represent a fixed distance, but was rather chosen to exclude transient
emission that was clearly remote from the main flare ribbons.

The resulting source area for each event is therefore the intersection
of three sets: the ribbon mask; near-PIL regions; and pixels within
the proximity threshold.  
We refer to pixels in this intersection as RPP pixels, for ribbon,
PIL, and proximity.  Since there are 26 masks for each
event, there are in principle 26 sets of RPP pixels for each event.
We note, however, that some choices of ribbon identification parameters
lead to no overlap between our PIL masks and the ribbon masks.
For this reason, there is only one RPP pixel in events \#6 and \#8.

We note that the decay index of the field above an individual pixel
might not accurately represent the rate of decay of the field over the
larger area through which an eruption must propagate.  Consequently,
increasing the number of pixels used to characterize the source
region's decay index might improve the correlation with CME speed.
Set against this, however, including fields drawn from a larger area
might introduce field structure that is irrelevant to the dynamics of
some eruption(s), which could diminish the speed versus decay index
correlation.  
For instance, while we have identified all PIL regions, not all PILs
participate in every eruption, so including the field structure above
only those PILs near ribbons seems appropriate.
As a practical matter, though, more pixels could be included easily
enough, by separately dilating the ribbon and PIL masks (the ``R'' and
first ``P'', respectively, in ``RPP''), which would likely increase the
number of pixels in the intersection.  In addition, the RPP
intersection itself could be further dilated.  In a case study, below,
we do briefly investigate the sensitivity of the estimated decay index
for one event to choices of masking parameters, but we leave a more
thorough investigation of modifications to our approach to a future
study

To model the fields overlying a CME's source area, we extrapolate the
potential magnetic field's components above each RPP pixel.
Like \citet{Xu2012}, we use a Cartesian, Green's function model.  This
approach treats the flux in each magnetogram pixel as a point source
({\it e.g.}, \citealt{Sakurai1982}), and assumes no sources exist
outside of the magnetogram.  The magnetograms are centered on the
source ARs, and most are a few hundred seconds of arc on a side,
meaning strong fields within several tens of megameters of RPP pixels
are included in the extrapolations.

No currents are present in our model field, but they are 
certainly present in the actual magnetic field in CME source regions. 
We expect, however, that current densities should usually decay
rapidly with height, and, further, that the potential field
approximates the actual field's strength, if not its orientation.
%
%

Next, we analyze the extrapolation(s) to characterize the overlying
field's structure.
Following \citet{Liu2008}, we characterize the rate of decay of the
overlying magnetic field with height by fitting the extrapolated
horizontal magnetic field strength, $B_{\rm h}  = \sqrt{B_x^2 + B_y^2}$, as a
function of radial height $r$.  For this fit, we assume $B_{\rm h}$ follows
a power law in $r$,
\be B_{\rm h} \propto r^{\gamma} ~, \label{eqn:powerlaw} \ee
where $\gamma$ is the decay index.
We fit $B_{\rm h}(r)$ for $r \in \{46.5, 59.2, 72.3, 85.8, 99.9, 114.5\}$
Mm.  
We have not investigated the orientation of the extrapolated field
over this range, so the field morphology might not be arcade-like, as
we assumed.

This range of heights matches that fitted by \citet{Liu2008} in
global, potential-field source-surface (PFSS) extrapolations.  These
particular fitted heights, however, match those used by
\citet{Suzuki2012} in fitting $\log(R_\odot + r)$, in which they are nearly
uniformly spaced.
Like \citet{Xu2012}, however, we used Green's function extrapolations
instead of PFSS models.
As noted above, \citet{Xu2012} fitted the logarithm of points linearly
spaced in height $r$, which biased their fitted decay indices toward
the decay rate at the large-$r$ end of their fitting interval.
Unfortunately, this bias from uneven sampling in $\log(r)$ did not
occur to us until after much of our study had been completed.  In
fact, the heights we fit are not evenly spaced in either $r$ or
$\log(r)$; while they are more evenly spaced in $\log(r)$ than points
evenly spaced in $r$ would be, we still over sample the high-$\log(r)$
end of the range compared to the low-$\log(r)$ end.
This can slightly weight our estimates of $\gamma$ by the rate of
decay for larger $r$, but since our focus is on variations in
$\gamma$ between events, a uniform bias should not be problematic for
our analysis.

We fit the horizontal magnetic field strength because we expect is a
better measure of tension from the overlying field than the total
field strength.
A more negative value of $\gamma$ corresponds to a more rapid decline
in $B_{\rm h}$ as a function of height and thus a weaker overlying field. In
our qualitative model, this would indicate that there is less downward
magnetic tension due to the overlying field, and we would expect the
CME to be faster than one from a source region with a less rapid
fall-off in $B_{\rm h}$.

\citet{Suzuki2012} used two methods to find $\gamma$, the fitted decay
rate, from $B_{\rm h}(r)$.  We use similar approaches here, estimating
$\gamma$ two ways for each event and set of RPP pixels.
In our Method 1, we first compute the average
of transverse field strengths over the RPP pixels at each height, and
then fit $\gamma$ from these average field values.
In our Method 2, we first fit $\gamma$ from $B_{\rm h}(r)$ for each RPP
pixel, and then average these $\gamma$'s to find a whole-RPP
$\gamma$. 
%
%
Comparisons of $\gamma$'s computed two ways enables estimating the
uncertainty in $\gamma.$
The results computed with each method are presented in Table
\ref{tab:data}; the significance of differences in these results will
be discussed in Section \ref{sec:results}. In both approaches, we used
a linear, least-squares fit to the logarithms of both horizontal field
strength, $B_{\rm h}$, and radial heights, $r$. For Method 1, we passed in
the standard deviation of $B_{\rm h}$ over the RPP pixels at each height to
the linear fitting procedure as an uncertainty weighting, to account
for varying field strengths at that height. The uncertainty in
$\gamma$ derived this way comes from the standard error of the
estimation procedure. For Method 2, the uncertainty of $\gamma$ comes
from the standard deviation of the $\gamma$ values computed from all
different pixels in the extrapolation region.

In our conceptual model of the CME process (an extension of the CSHKP
cartoon), reconnection of magnetic fields beneath the erupting
structure might actively contribute to driving the CME's eruption.
This might occur either: directly, {\it e.g.}, by momentum transfer to the
CME from the ``dipolarizing,'' post-reconnection magnetic fields; or
indirectly, {\it e.g.}, by increasing the amount of flux that wraps around
the ejection, thereby increasing the hoop force on the ejection.
This reconnection can be measured by total photospheric flux swept out
by flare ribbons. We calculated the magnetic flux by first taking the
absolute values of the magnetograms and then summing over the product
of these unsigned field strengths and the ribbon mask.
Note that this double counts the reconnected magnetic flux; but since
our analysis focuses on linear correlations, this factor of two is
irrelevant.
Since for each event we have 26 different ribbon masks, we used the
average of the fluxes calculated from each of the masks and estimated
the uncertainty of the flux using the standard deviation of these
fluxes. The fluxes computed this way are listed in Table
\ref{tab:data}.
%
%
%
%

\subsection{Case Study: A Typical Event} 
\label{subsec:case}

In this section, we examine in detail the event on 26 July 2002 (\#
9), to illustrate how selection of RPP pixels and modeling of the
overlying field was done, and to study the effect of different methods
and parameters on our calculated value of $\gamma$. The CME associated
with this event had a speed of 818 km s$^{-1}$ and ribbon flux 2.6
$\times 10^{21}$ Mx.  These are both around the average levels in our
event list (although this speed is fast compared to most CMEs; {\it e.g.},
\citealt{Gopalswamy2010}).  We chose this event as a typical example
of those in our list of CMEs to investigate the validity and
robustness of our basic approach, as opposed to any particular
features of this event. The magnetogram of this event exhibits some
artifacts that are present in some other events' magnetograms.
We developed different extrapolation methods to test how different our
results could be and how robust our original method is to these
imperfections of our input data.

\begin{figure}
  \centerline{\includegraphics[width=8cm,clip]{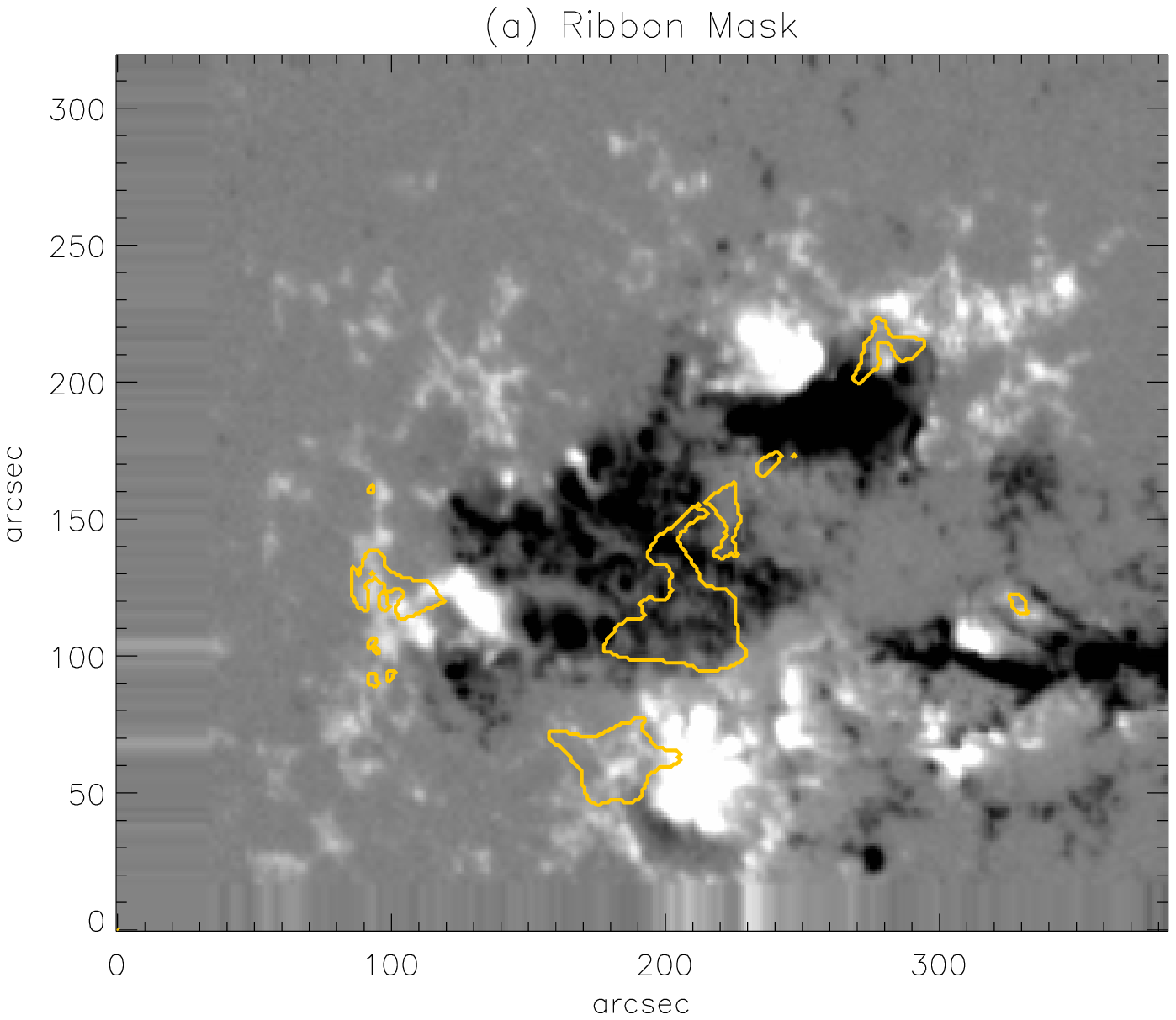} %
    \includegraphics[width=8cm,clip]{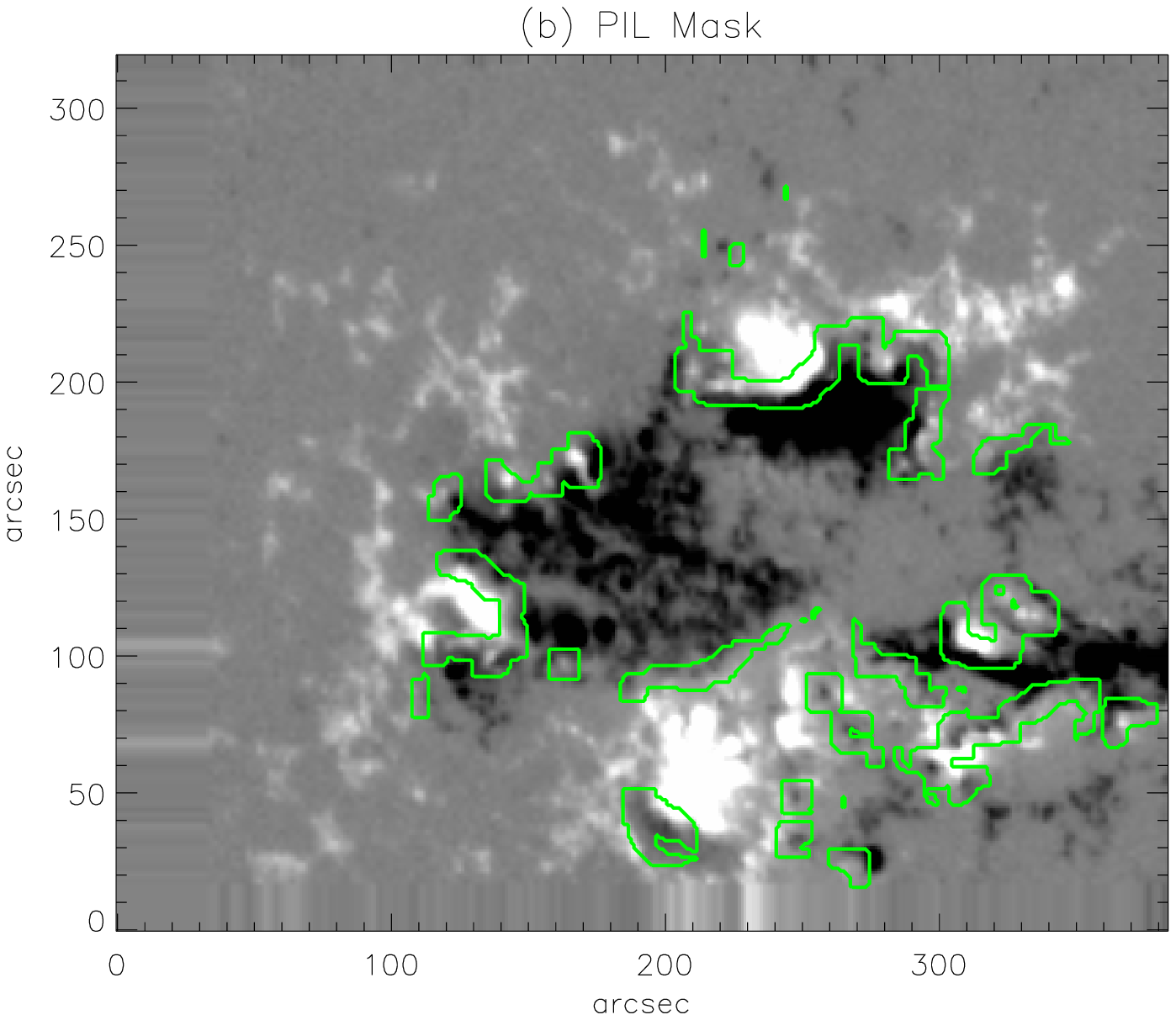}}
 \caption{(a) Contours of derived ribbon mask overlain on the
   interpolated full-disk MDI magnetogram for event \# 9.  (b)
   Contours of derived PIL regions over the same magnetogram. The
   magnetograms are displayed with a saturation level of 500 Mx
   cm$^{-2}$, and the white areas represent positive fields whereas
   black areas represent negative fields.}
 \label{fig:ribbonmask&pilmask}
\end{figure}
\begin{figure}
 \centerline{ \includegraphics[width=8cm,clip]{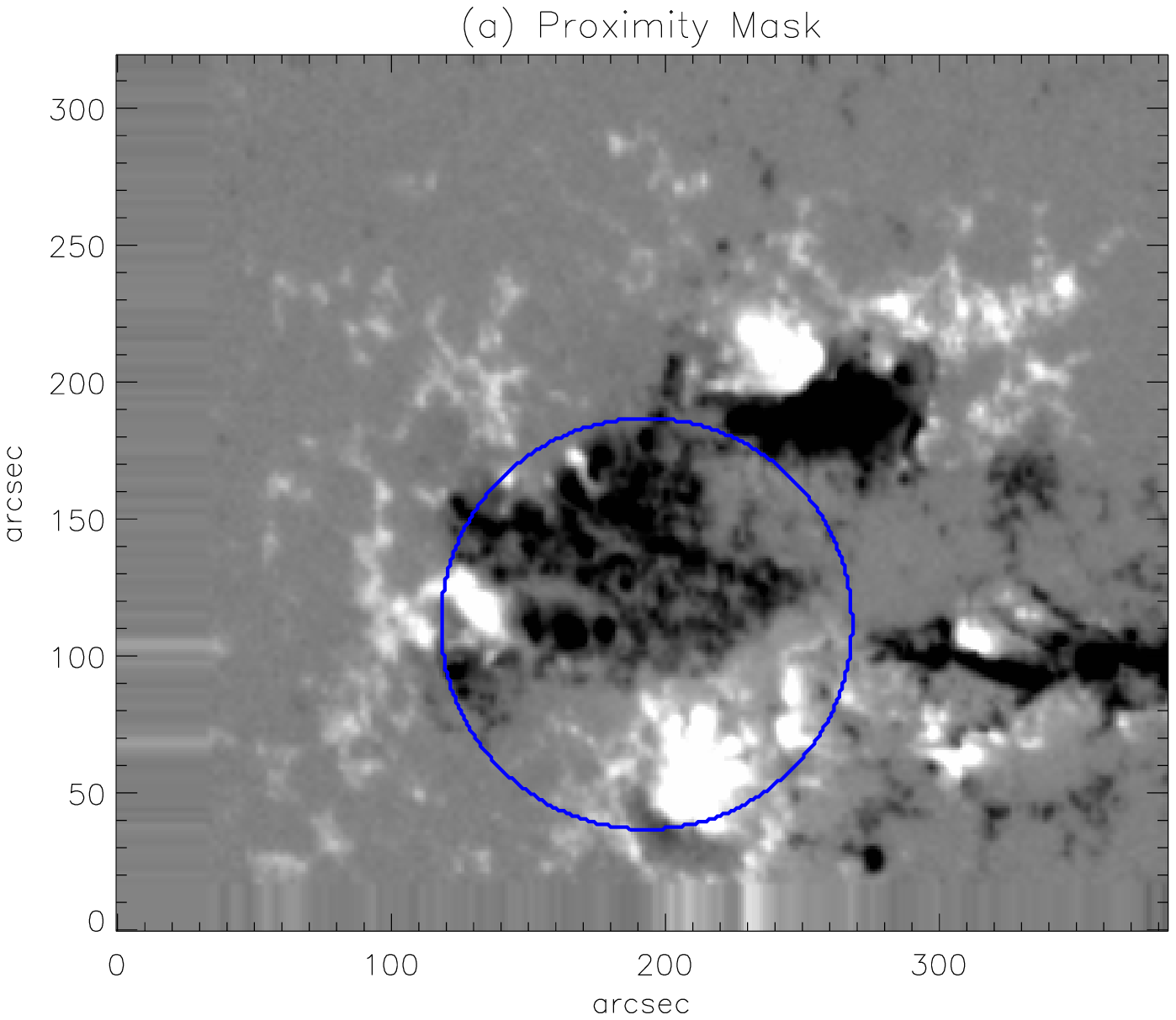} %
  \includegraphics[width=8cm,clip]{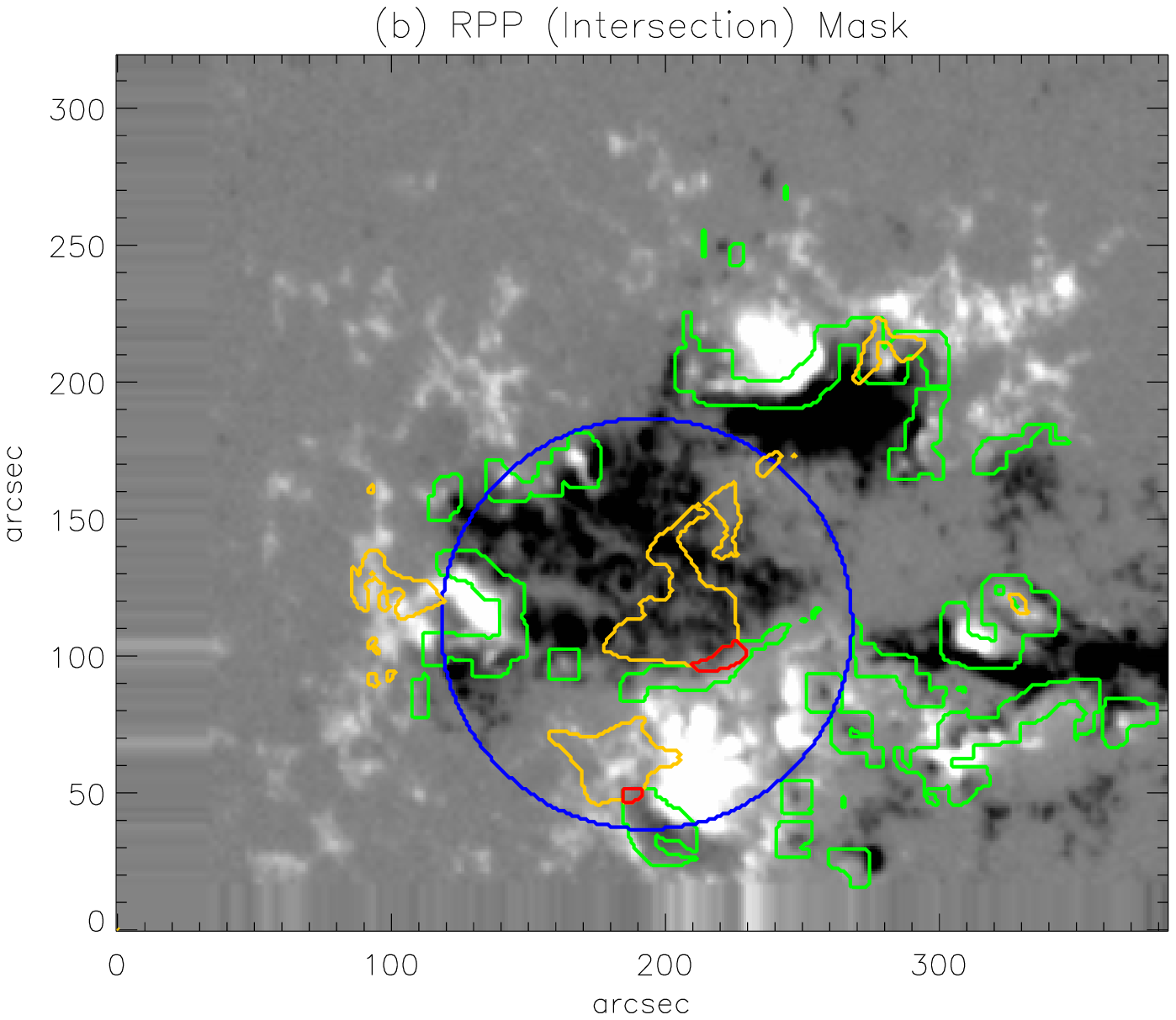}}
 \caption{(a) Contours of a circular mask of radius 75 pixels
   centered at the mean position of ribbon pixels. (b) RPP
   pixels (red) lie in the intersection of ribbon mask (yellow), PIL
   regions (green), and the circular mask (blue). The PIL region is
   selected with a threshold of 100 Mx cm$^{-2}$. The potential field
   extrapolation was carried out above all red pixels.}
 \label{fig:circle&intersection}
\end{figure}

The flare ribbon locations for this event, found from H$\alpha$ flare
observations, are shown (yellow contours) in Figure
\ref{fig:ribbonmask&pilmask}a, plotted over the corresponding
magnetogram.  In Figure \ref{fig:ribbonmask&pilmask}b, we
show PIL regions (green contours) identified from the magnetogram.  In
Figure \ref{fig:circle&intersection}a, we show the 75-pixel proximity
threshold (blue circle) centered on the mean position of the ribbon
mask for this event.  In Figure \ref{fig:circle&intersection}b, we
show RPP pixels (red contours), which are the intersection of all
three sets.  Note that not all intersections of PIL regions and the
ribbon mask are confined to a small area; some intersections are
relatively far from the center of the ribbon emission, {\it e.g.}, the
relatively small patches of yellow at $(x,y)$=(275,215) and (325,120).
As noted above, we do not expect the magnetic field directly above
such small-scale, remote areas of ribbon emission to significantly
influence the dynamics of the erupting CME, so they are excluded from
our analysis by the proximity threshold.  Accordingly, we only include
RPP pixels in the potential field extrapolation.

%

%

Having defined the source region for this event, the extrapolation can
be carried out. Using a Green's function method, horizontal magnetic field
strengths at points above the selected pixels were calculated from the
line of sight field strengths of all other points in the
magnetogram. This is done at heights of $\{46.5, 59.2, 72.3, 85.8,
99.9, 114.5\}$ Mm. According to our assumed power-law relation between
CME speeds and the transverse field strengths, the power-law index
$\gamma$ is $-2.26 \pm 0.2$  using Method 1 and $-2.09\pm 0.3$
using Method 2.

How robust is our estimate of this decay index?  To find out, we
  conducted several tests, which we outline in the following
  paragraphs.

\begin{figure}
 \centerline{ \includegraphics[width=8cm,clip]{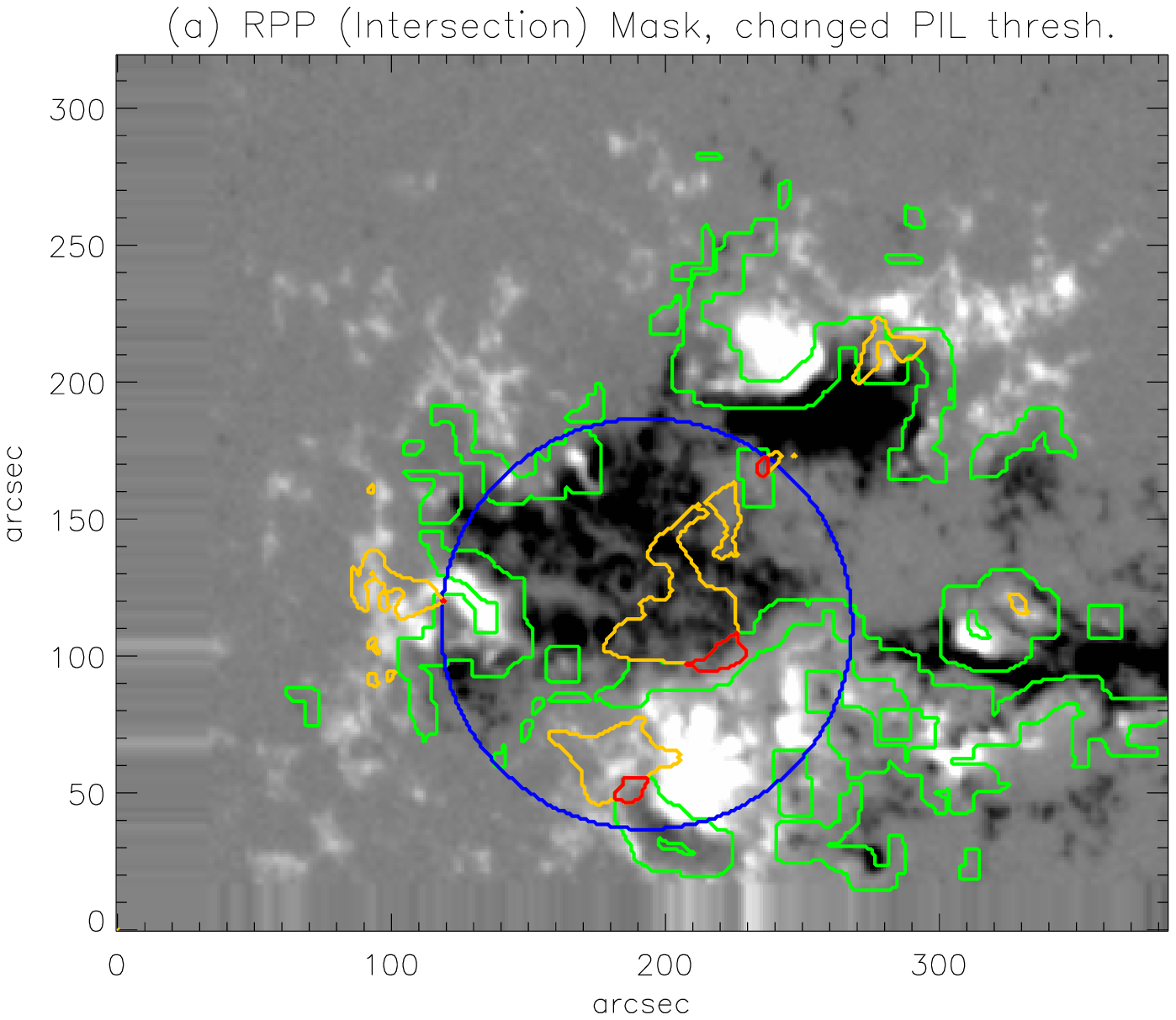} %
  \includegraphics[width=8cm,clip]{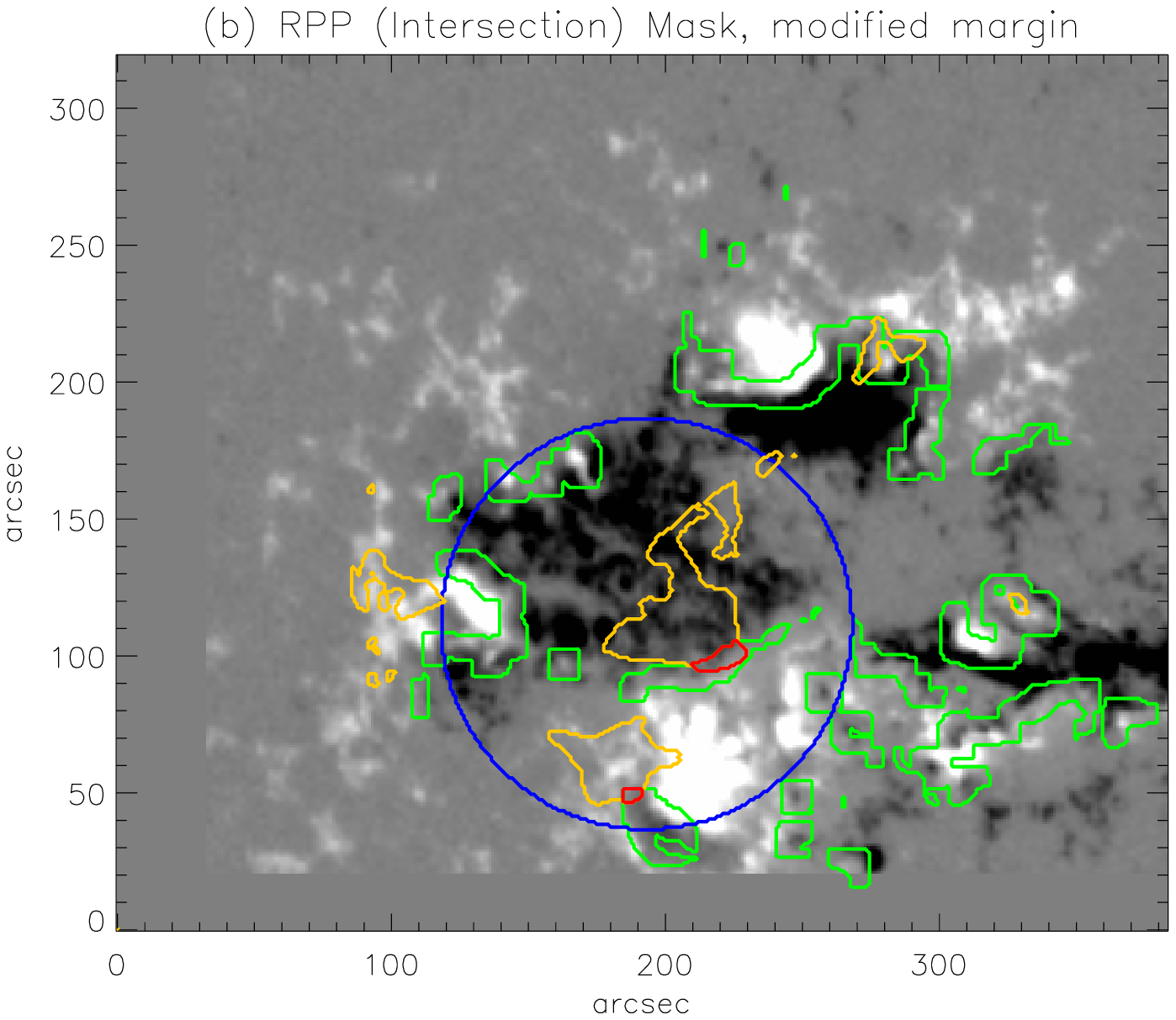}}
 \caption{(a) We show ribbon pixels (yellow), the proximity
   threshold (blue), PILs regions (green), and RPP pixels (red), but
   with the PIL regions determined using a different threshold: 50 Mx
   cm$^{-2}$ instead of the 100 Mx cm$^{-2}$ used in 
   Figure \ref{fig:circle&intersection}b.  Comparison with that figure
   shows that more RPP pixels are identified, including a new, small
   patch separated from the other patches. (b) The same quantities,
   but with the original PIL threshold, and the magnetic field
   artifacts at the magnetogram margins set to zero.  Changes in
   neither the PIL threshold nor magnetogram margins significantly
   affect $\gamma$.}
 \label{fig:diffpil&nomargin}
\end{figure}

To start, we tried some different parameters when deriving some of
the key components for extrapolation to test the robustness of our
methods. First, we note that when selecting the PILs, we used a rather
high threshold 100 Mx cm$^{-2}$ to exclude insignificant PILs. This
substantially reduces the number of pixels identified for potential
field calculation. To see if including more pixels would alter the
result, we lowered the threshold to 50 Mx cm$^{-2}$. 
Figure \ref{fig:diffpil&nomargin}a shows the new PIL areas selected,
which encompass many more pixels than with the higher threshold (Figure \ref {fig:ribbonmask&pilmask}b). Following the same
extrapolation and fitting procedures, we determined the decay index
from Method 1 to be -2.15, which is slightly lower than what we found
with fewer pixels. This is expected since we now include some
peripheral pixels with weaker fields. Still, this result is within one
standard deviation of the original, indicating that original approach
is robust.

Also, we note that there are bad margins present in magnetograms for
some of the events, including both this example and events $\#1$,
$\#3$, $\#4$, $\#5$, $\#11$, $\#12$, $\#15$ and $\#16$.  In these
cases, field values are replicated from the edge of the estimated
chromospheric field values to the margins of the magnetogram arrays.
How significantly do these artifacts affect our estimated decay
indices?
To investigate the effect of bad margins on our decay index estimate
for this particular event, we excluded the artifacts by setting the
field there to zero (Figure \ref{fig:diffpil&nomargin}b). The
resulting $\gamma$ from Method 1 based on this modified magnetogram is
-2.31 which is slightly higher than originally computed but also
within one standard deviation.

It is possible that the limited fields of view of our
  magnetograms could also affect our estimated decay indices, since
  fields outside of our magnetograms are ignored.  But even within a
  magnetogram, how sensitively do our estimated decay indices depend
  upon magnetic flux relatively far from the flaring region?  To see
if our use of limited information matters, we decided to restrict our
information to a circle with radius 150 pixels at the
weighted event center. This specific radius was chosen because in the
extrapolation for this event, the range of the heights are roughly
comparable to 150 pixels and it is believed that, for the Green's
function extrapolation of a localized source distribution, the main
contribution of the field at height $H$ comes from fields with a
circle of radius $H$ centered at that point. A new test magnetogram was
then obtained by setting the line-of-sight field outside the circle to
zero. For this test, the decay index computed by Method 1 is -2.17, which
is again only slightly different from our original value. This
suggests that relatively distant magnetic flux should not
significantly alter our estimates.  These tests suggest that our
original approach for estimating $\gamma$ values do not depend
sensitively on our parameter choices or details of the magnetograms
that we analyze.


\section{Results}
\label{sec:results}

\begin{figure}
 \centerline{\includegraphics[width=8cm,clip]{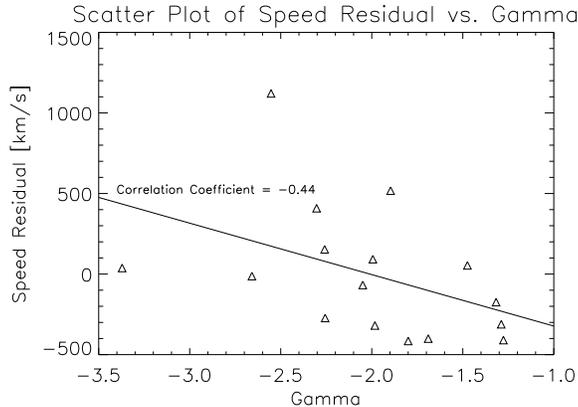}}
 \caption{Scatter plot of CME speeds' residual dependence on $\gamma$,
   after removal of the dependence of both speeds and $\gamma$ upon
   ribbon flux $\psi$.}
 \label{fig:scatter_residual}
\end{figure}

\begin{figure}
 \centerline{\includegraphics[width=8cm,clip]{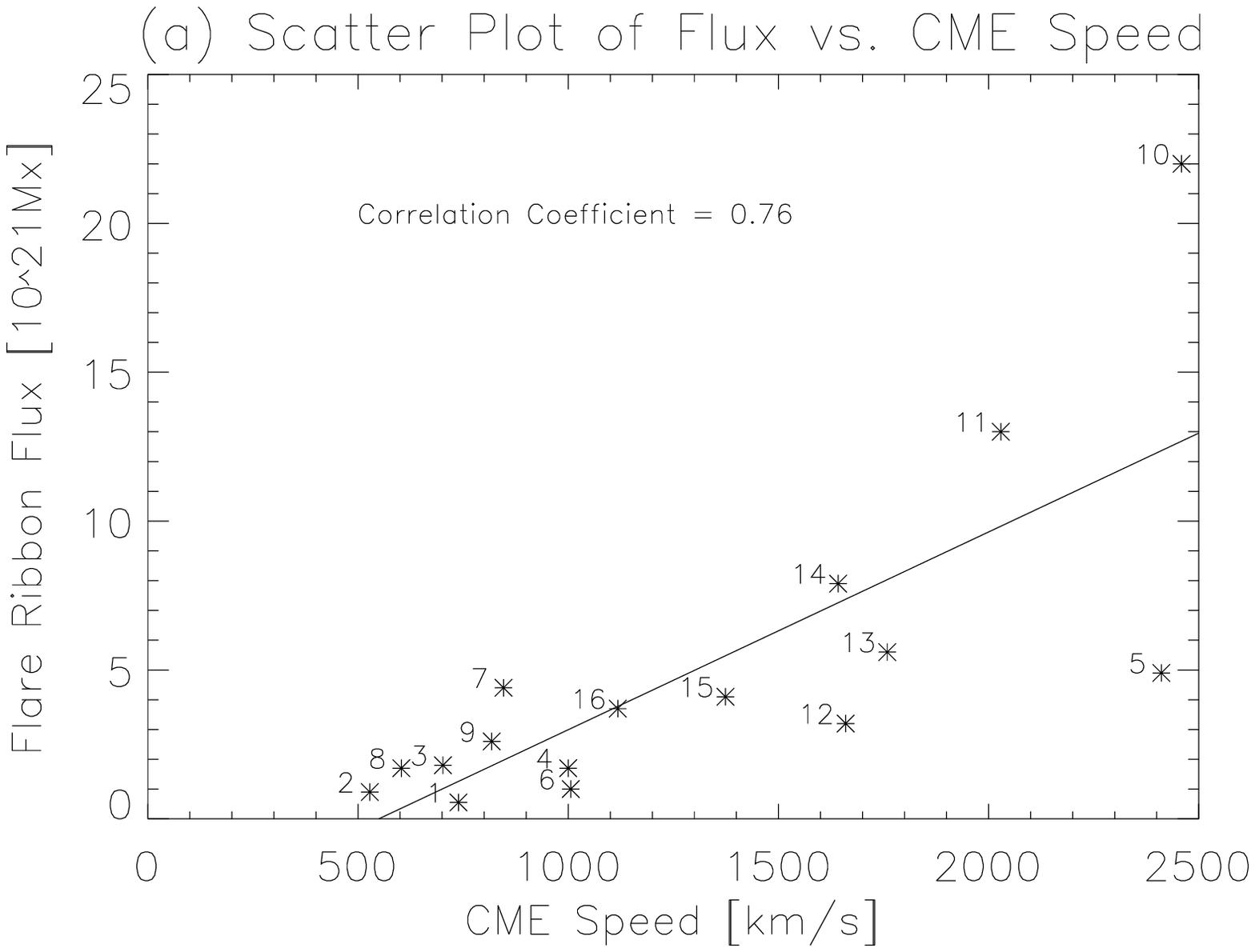} %
  \includegraphics[width=8cm,clip]{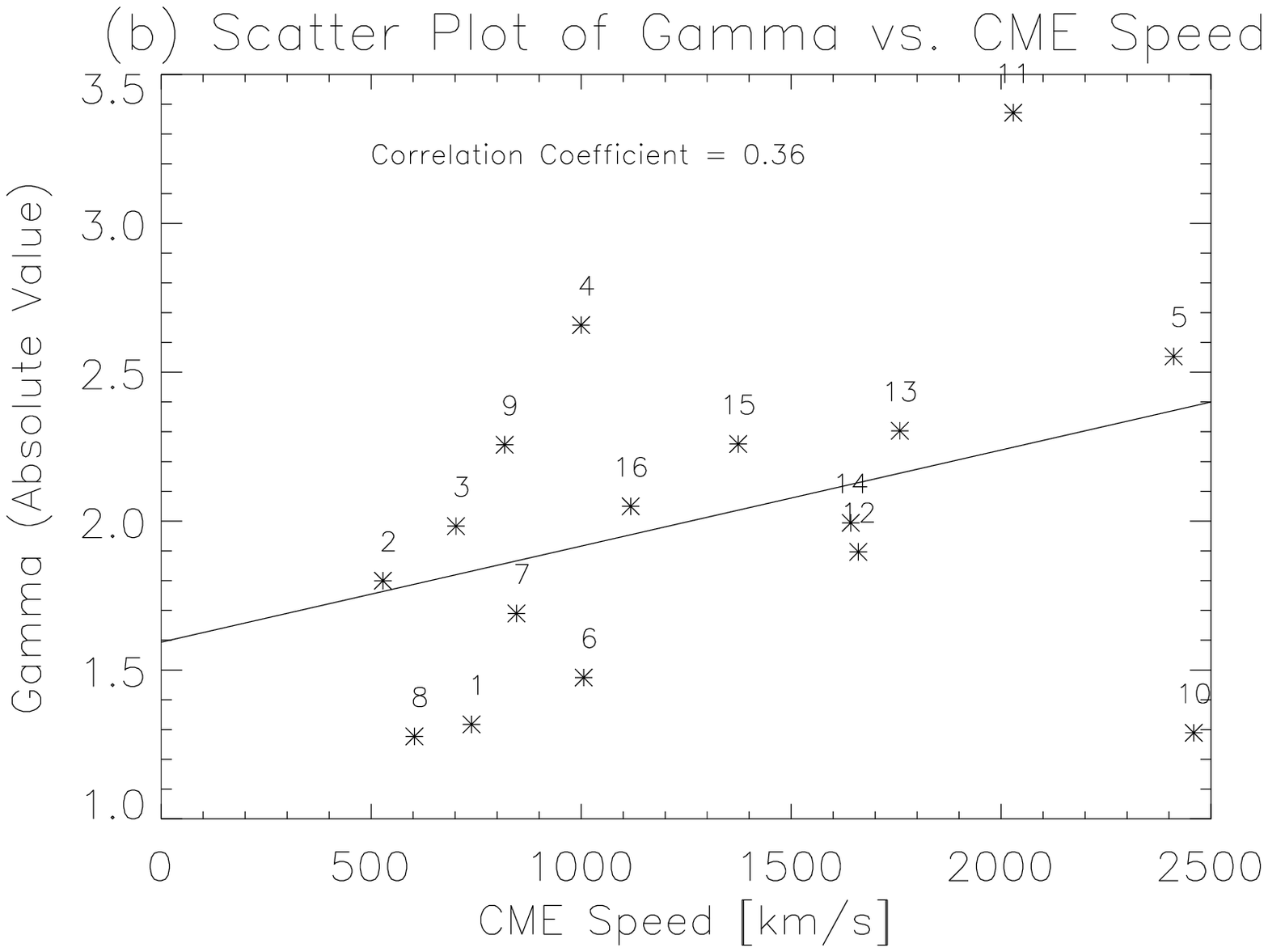}}
 \caption{(a) Ribbon fluxes as a function of CME speeds. (b)
   Gamma as a function of CME speeds.  Event numbers are shown with
   each plotted symbol.}
 \label{fig:reverseplot}
\end{figure}

Following the approach discussed above, we carried out potential field
extrapolation for all 16 events, for all masks with nonzero ribbon
pixels. We then computed the decay index based on these
extrapolations. The speed of each CME, $v_{\rm CME}$, was taken from
the CDAW CME list.  A significant correlation between CME speeds
($v_{\rm CME}$), ribbon fluxes ($\psi$), and decay indices ($\gamma$)
was found. The CME speed data and ribbon flux yield a linear
correlation coefficient of 0.76, with coefficient of determination
0.58; see Figure \ref{fig:reverseplot}a.
The scatter plot of $v_{\rm CME}$ versus $\gamma$ in
Figure \ref{fig:reverseplot}b also shows a clear
correlation, with a linear correlation coefficient of -0.36, and
coefficient of determination 0.13.
A negative correlation between $\gamma$ --- in which more negative
values correspond to more rapidly decaying fields --- and CME speeds
accords with stronger overlying fields exerting stronger downward
tension forces that lower the speeds of CMEs.
To account for these mutual correlations, we compute the partial
correlation between CME speed residuals and $\gamma$, which excludes
the effect of flux $\psi$ in both variables.  This yields a
correlation of -0.44 with coefficient of determination 0.19; see
Figure \ref{fig:scatter_residual}. The existence of a residual
correlation implies that extra information about CME speeds is
contained in the decay index that is not accounted for by the ribbon
flux.  So, while it is plausible that $\gamma$ and $\psi$ might be
physically related ({\it i.e.}, not physically independent variables),
neither is redundant.

Accordinly, we must characterize the joint dependence of $v_{\rm
  CME}$ on $\gamma$ and $\psi$.
To do so, we fitted the CME speeds with a two-variable linear
regression on $\psi$ and $\gamma$, with all three variables
standardized ({\it i.e.}, the mean of each was removed, and the result was
scaled by its standard deviation). The linear regression line was
calculated to be
\be v_{\rm CME}=0.73\psi-0.29\gamma ~, \label{eqn:coeffs} \ee 
where $v_{\rm CME}$ is the standardized speed and $\psi$ is the
standardized magnetic flux.  
(Note that our use of standardized variables here implies that no
constant term is fitted in the model.)
The multiple correlation coefficient ---
the correlation between the predicted and actual speeds --- is 0.81,
with a coefficient of determination of 0.66.  For linearly related
variables with Gaussian deviations, a coefficient of determination
equal to one would imply that the dependent variable is completely
determined by the independent variables; the difference between this
coefficient and one is the fraction of variance unexplained by the
independent variable(s).
Under these assumptions, the increase in coefficient of determination
from 0.58 for $\psi$ alone to 0.66 for both $\psi$ and $\gamma$
represents a roughly 20\% decrease in unexplained variance.
Clearly, even with both variables, a substantial fraction of the
variability in CME speeds remains unaccounted for.

To characterize the sensitivity of our results to the particular set
of 16 events that comprise our sample, we estimated the variances in
our results using a delete-1 jackknife procedure \citep{Efron1982},
{\it i.e.}, by running our regression analysis on each of 16 subsets created
from our event sample, each with one event deleted.
%
%
We found the mean coefficient of $\psi$ in our regression to be 0.736
with standard deviation 0.034, and the mean coefficient of $\gamma$ to
be -0.293 with standard deviation 0.062.  Applying the delete-1
procedure to our standardized variables implies the means of each
variable in the resulting 15-event subsets depart from zero, {\it i.e.}, 
the subsets are no longer standardized.  In this case, our regression 
procedure also generally returns a nonzero constant term.   The mean
and standard deviation of the fitted constant terms are 
0.0036 and 0.045, respectively.
Finally, we found the mean and standard deviation of the coefficient
of determination to be 0.66 and 0.045, respectively.

Our use of standardized variables implies that the independent
variables in Equation (\ref{eqn:coeffs}) enter with equal weight,
implying their predictive power can be directly compared from their
regression coefficients.
Consequently, the larger coefficient for $\psi$, 0.73 versus 0.29 for
$\gamma$, implies that ribbon fluxes are much stronger predictors of
CME speeds than decay indices.



\begin{figure}
 \centerline{\includegraphics[width=8cm,clip]{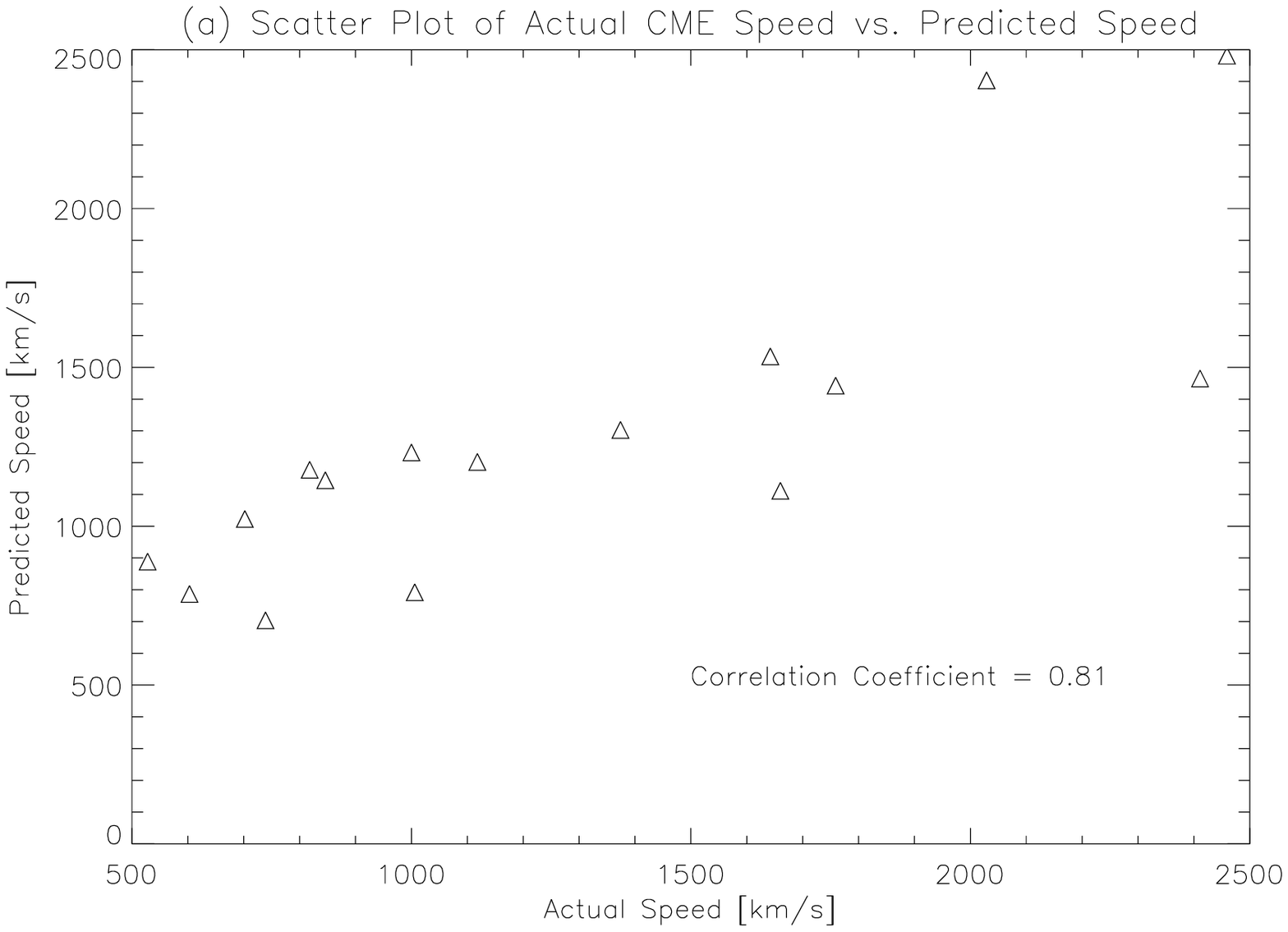} %
  \includegraphics[width=8cm,clip]{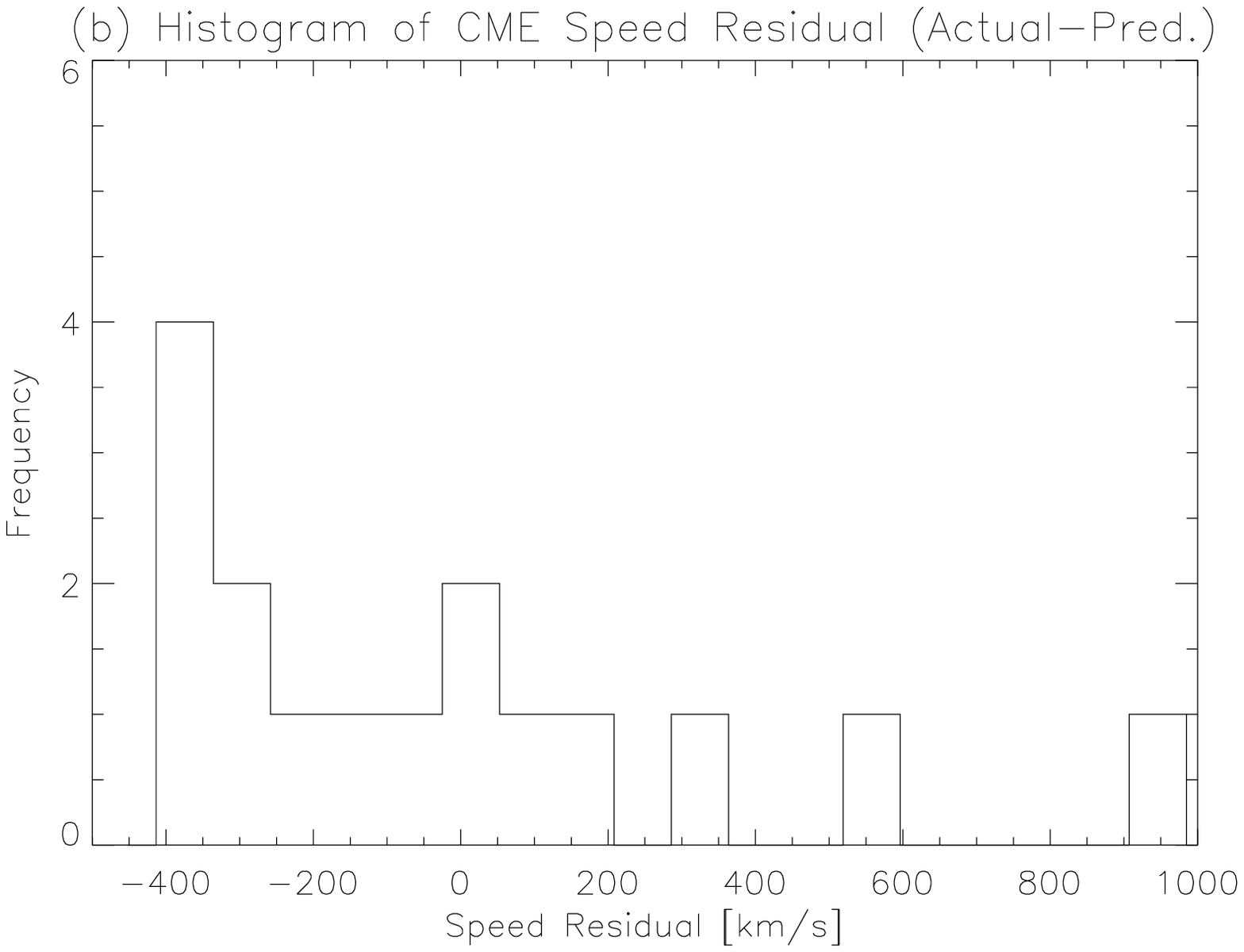}}
 \caption{(a) A scatter plot of actual CME speeds {\it vs.} predicted
   values. (b) Histogram of deviations of actual CME speeds from
   the linear regression predictions, in original units.}
 \label{fig:scatter&hist}
\end{figure}

In Figure \ref{fig:scatter&hist}a, we plot the
predicted CME speeds, based upon the regression in Equation
(\ref{eqn:coeffs}), versus the actual CME speeds.
The histogram of residuals between the actual CME speeds and the
linear regression predictions is plotted in Figure \ref{fig:scatter&hist}b,
computed in physical units.  
Discrepancies tend to be larger for very fast CMEs, compensated by a 
slight tendency for the model to overestimate speeds: there are 9
overestimates versus 7 underestimates.
The outlier at the right end corresponds to event number 5.
It is plausible that CME dynamics depend non-linearly on the variables
we consider, instead of the linear dependence that we assume; but
discriminating between the linear model and a more complex model would
require a larger event sample.

Also, inspired by the work of \citet{Robbrecht2009} on stealth CMEs
---CMEs observed without ``low coronal signatures'' --- we plotted
flux and gamma as a function of CME speed to see if there 
  were some lower bounds for flux and decay index below which
there is no eruption at all. The flux versus CME speeds plot in
Figure \ref{fig:reverseplot}a suggests that there is no minimum ribbon flux
required for CMEs, a result expected given reports of stealth CMEs.
Usually such CMEs are slow ($\approx 300$ km s$^{-1}$; \citealt{Ma2010}),
and in order for a CME to be accelerated to a speed greater that the
ambient speed of the solar wind ($\approx 450$ km s$^{-1}$), a significant amount
of coronal magnetic reconnection, with consequent ribbon fluxes, appears
necessary. 
In contrast, our fit of speeds versus $\gamma$ implies that,
statistically, a minimum absolute value of $\gamma$ around 1.6 is
required for CMEs.  In our sample, though, several CMEs actually occur
even with values of $\gamma$ less than this, highlighting that the
fitted threshold is statistical.
%

\section{Discussion} 
\label{sec:discussion}

 
We have investigated the combined roles of (i) flare ribbon flux, $\psi$, and
(ii) the decay rate of overlying fields, $\gamma$, in a set of 16 eruptive
events. We have confirmed previous results that higher CME speeds are
associated with both larger ribbon fluxes and more rapidly decaying
overlying fields.  Comparing the two effects, we found that CME speeds
depend more strongly on ribbon fluxes than the rate of decay of
overlying fields.  

Using a simple linear regression, we found that these two factors
alone can explain 66\% of the variation in CME speeds.  Given existing
routine observations of (i) photospheric magnetic fields and (ii) EUV
and/or H$\alpha$ emission that can be used to identify flare ribbons,
our results suggest that observers can readily make an approximate
prediction of the speed of a flare-associated CME at the time of its
eruption, prior to its appearance in a coronagraph.
In terms of fractional errors in the predicted speeds of the
relatively fast, halo CMEs in our sample, our mean $|dv|/v$ is 25\%,
and our mean $dv/v$ is -10\%.  
Velocity estimates based on our results could be incorporated into
existing CME propagation models ({\it e.g.}, \citealt{Zhao2014}).
%
%
Independent of variations in CME speeds due to acceleration occurring
in interplanetary space, our errors in predicting CME speeds imply
mean signed and unsigned errors in CME arrival times of 4 h and 10
h, respectively.
This compares favorably with the current approach of fitting
displacements of CME structure in LASCO image sequences ({\it e.g.},
\citealt{Mays2015}), although we have only analyzed a small sample.
%
%

We note that, compared to the study by \citet{Qiu2005}, we included some
additional events.  In our study, we found that the correlation of CME
speeds with ribbon fluxes was weaker than the correlation they
reported.  We hypothesize that either the effect of outliers in their
smaller sample or new events within our larger sample are
responsible for the difference.

We remark that $\gamma$ values might depend significantly upon the
method used to determine the potential field.  Before we extrapolated
fields using the Green's function method, we first tried using PFSS
models to estimate $\gamma$ above active regions in a sample of 47
CMEs. Following essentially the same PIL identification method
discussed above (but without ribbon observations to identify which PIL
regions were involved in these flares), we calculated the decay index
$\gamma$ above PILs in these regions. In Table \ref{tab:gamma_comp},
we compare our results from the Green's function method used described
above with $\gamma$'s estimated from PFSS models.  We find that Green's
function $\gamma$'s are systematically more negative than PFSS
$\gamma$s, implying more rapid decrease of horizontal field strength
with height.

Given the similar approaches used, why do the inferred values for
  $\gamma$ differ systematically?  First, it should be noted that
these two sets of extrapolations were done above different sets
pixels, which might by itself explain the differing $\gamma$ values.
But the difference might also be due to the lower spatial resolution
of typical PFSS models (ours used spherical harmonics up to $\ell =$
192), which yield smaller field strengths at low coronal heights.
While our Green's function method treats each pixel in the magnetogram
as a single, delta-function source, PFSS models typically integrate
over pixels to compute spherical harmonic coefficients. In this way,
small bipoles (positive and negative flux close to each other) can
get averaged out in low-resolution PFSS models, thus not contributing
to the extrapolated field.  With their better resolution, however,
Green's function methods can incorporate fields from small-scale
bipoles into the extrapolations.  Since the dipole moments from these
small-scale fields are quite strong just above the photosphere but
decay relatively rapidly with height, the field strengths computed
using Green's functions probably systematically decrease faster with
height than PFSS fields, resulting in a more negative $\gamma$.
A more detailed study comparing methods of estimating decay indices
would be worthwhile.

Why do CME speeds depend more strongly on ribbon fluxes than decay
indices?  One possibility is that this is an artifact of our approach:
since we only studied events with successful CME onset, we did not
take into account the fact that there might be failed eruptions due to
downward tension caused by overlying fields ({\it e.g.}, \citealt{Wang2007b,
  Liu2008}).
In this sense, part of the effect of the decay rate of the overlying
field on a CME eruption is obscured compared to the effect of ribbon
fluxes: $\gamma$ might need to be negative enough for an eruption to
be permitted at all, {\it i.e.}, a high enough decay rate might be a
necessary condition for an eruption in the first place.
(Given that \citet{Xu2012} report eruptions for a very wide range of
$\gamma$ values, however, it is not clear if any meaningful threshold
in $\gamma$ exists.)
Another possiblity is that potential field models are too inaccurate:
perhaps discrepancies between the potential and actual magnetic fields
above eruption sites are significant enough that our estimates for
decay indices are wrong.  Or perhaps decay indices themselves are
irrelevant: maybe lateral motions in the early stages of eruptions
enable CMEs to erupt ``sideways'' rather than pushing through
overlying fields as we assume.

Ultimately, it is clear that factors other than just the two that we
consider also govern CME speeds.
Likely possibilities include the internal magnetic
structure of the pre-flare magnetic system that later erupts ({\it e.g.},
the amount of flux within this system, and / or the degree of its
initial magnetic twist, perhaps quantified by magnetic helicity) and
interplanetary magnetic field and density structures (including the
effects of previous CMEs).
%
%
We also remark that the CME speeds that we cite are plane-of-sky
speeds; CMEs' true speeds might be faster, which is an additional
source of discrepancy between our simplistic model and the observations.

\begin{table}
\footnotesize
\label{tab:data}
\caption{Under CME Label, data sources are identified by
  H(H$\alpha$) and T(TRACE). We note that compared to \citet{Qiu2005}
  and \citet{Qiu2007}, we add event $\#1$ and $\#2$. The information
  about date, time and speed of these events are from CDAW CME list,
  and the information about location and AR number are from UH IfA
  active region label maps. 
  TRACE's $0.''5$ pixels were summed 2 x 2.
  The * in the Decay indices (Method 2) of events
  $\#6$ and $\#8$ denotes that in the calculation of $\gamma$ for the
  two events, there is only one pixel in the RPP regions,
  thus precluding estimating standard deviations.}

\begin{tabular}{llllllllll}
\hline
CME & Date & Time & Location & $v_{\rm CME}$ & AR & Gamma & Gamma & Flux & Pixel\\
Label &  & [hh:mm:ss] &    & [km s$^{-1}$]  & \# & (Method 1) & (Method 2) & [10$^{21}$Mx] & size ($''$)\\
\hline
1(H) & 2000 Jan 18 & 16:03:02 & S18E14 & 739 & 8831 & $-1.32\pm0.1$ & $-1.28\pm0.1$ & $0.55\pm1.1$ & 1.16\\
2(T) & 2000 Jul 25 & 01:39:01 & N06W01 & 528 & 9097 & $-1.80\pm0.4$ & $-1.79\pm0.3$ & $0.9\pm0.2$ & 1\\
3(H) & 2000 Aug 09 & 14:23:01 & N11W01 & 702 & 9114 & $-1.98\pm0.1$ & $-2.07\pm0.1$ & $1.8\pm1.5$ & 1.05\\
4(H) & 2000 Nov 24 & 20:48:02 & N21E06 & 1000 & 9236 & $-2.66\pm0.2$ & $-3.12\pm0.5$ & $1.7\pm0.3$ & 1.12 \\
5(T) & 2001 Apr 10 & 04:48:02 & S22W07 & 2411 & 9415 & $-2.55\pm0.1$ & $-2.19\pm0.1$ & $4.9\pm0.5$ & 1\\
6(T) & 2001 Apr 26 & 09:36:03 & N16W15 & 1006 & 9433 & $-1.47\pm0.2$ & $-1.47 *$ & $1\pm1.4$ & 1\\
7(H) & 2001 Sep 28 & 07:59:02 & N13E26 & 846 & 9636 & $-1.69\pm0.1$ & $-1.75\pm0.1$ & $4.4\pm0.6$ & 1.07\\
8(H) & 2002 Mar 20 & 16:00:01 & S19W06 & 603 & 9871 & $-1.28\pm0.1$ & $-1.30 *$ & $1.7\pm0.3$ & 1.07\\
9(H) & 2002 Jul 26 & 19:11:00 & S16E34 & 818 & 10039 & $-2.26\pm0.2$ & $-2.09\pm0.3$ & $2.6\pm1.3$ & 1\\
10(T) & 2003 Oct 28 & 09:35:03 & S16E18 & 2459 & 10486 & $-1.29\pm1.3$ & $-0.79\pm1.0$ & $22\pm3.2$ & 1\\
11(T) & 2003 Oct 29 & 19:11:03 & S17E04 & 2029 & 10486 & $-3.37\pm0.3$ & $-3.32\pm0.4$ & $13\pm1.2$ & 1\\
12(T) & 2003 Nov 18 & 06:23:03 & N03E35 & 1660 & 10501 & $-1.90\pm0.2$ & $-2.02\pm0.2$ & $3.2\pm0.9$ & 1\\
13(T) & 2004 Nov 07 & 14:23:03 & N09W08 & 1759 & 10696 & $-2.30\pm0.1$ & $-2.31\pm0.1$ & $5.6\pm0.7$ & 1\\
14(H) & 2005 May 13 & 16:03:02 & N12E19 & 1642 & 10759 & $-1.99\pm0.3$ & $-1.95\pm0.1$ & $7.9\pm0.5$ & 1.07\\
15(H) & 1998 Apr 29 & 15:59:04 & S17E30 & 1374 & 8210 & $-2.26\pm0.1$ & $-2.20\pm0.1$ & $4.1\pm0.5$ & 1.04\\
16(H) & 1998 Nov 05 & 15:36:04 & N18W07 & 1118 & 8375 & $-2.05\pm0.5$ & $-2.09\pm0.2$ & $3.7\pm0.6$ & 1.04
\end{tabular}
\end{table}

\begin{table}
\label{tab:gamma_comp}
\caption{In this table, we compare $\gamma$'s computed using
  Method 1 with our Green's function method with $\gamma$'s from PFSS
  models for an overlapping set of seven events. Apparently the
  magnitudes of $\gamma$'s computed from PFSS models are either about the
  same or smaller that $\gamma$'s computed with Method 1.} 
  
\begin{tabular}{lll}
\hline
Event & $\gamma$, & $\gamma$,\\
Number & Green's function & PFSS\\
\hline
1 & $-1.32 \pm 0.09$ & $-1.03 \pm 0.152$\\
4 & $-2.66 \pm 0.156$ & $-1.11 \pm 0.108$\\
5 & $-2.26 \pm 0.144$ & $-2.17 \pm 0.281$\\
6 & $-1.47 \pm 0.187$ & $-1.51 \pm 1.037$\\
10 & $-1.29 \pm 1.29$ & $-1.30 \pm 0.101$\\
11 & $-3.37 \pm 0.278$ & $-1.15 \pm 0.120$\\
14 & $-1.99 \pm 0.259$ & $-1.03 \pm 0.046$\\
\end{tabular}
\end{table}

\acknowledgments%
%
We thank Jiong Qiu
for kindly furnishing her flare ribbon data, and for insightful
comments that led us to improve the manuscript.
We thank the anonymous referee for constructive comments that improved
the manuscript.
We acknowledge funding from the NSF's National Space Weather Program
under award AGS-1024862, the NASA Heliophysics Theory Program (grant
NNX11AJ65G),  and the Coronal Global Evolutionary Model (CGEM) award
NSF AGS 1321474.
The CDAW CME catalog is generated and maintained at the CDAW Data
Center by NASA and The Catholic University of America in cooperation
with the Naval Research Laboratory. SOHO is a project of international
cooperation between ESA and NASA.


%

\begin{thebibliography}{27}
\ifx\bisbn     \undefined \def\bisbn  #1{ISBN #1}\fi
\ifx\binits    \undefined \def\binits#1{#1}\fi
\ifx\bauthor   \undefined \def\bauthor#1{#1}\fi
\ifx\batitle   \undefined \def\batitle#1{#1}\fi
\ifx\bjtitle   \undefined \def\bjtitle#1{\textit{#1}}\fi
\ifx\bvolume   \undefined \def\bvolume#1{\textbf{#1}}\fi
\ifx\byear     \undefined \def\byear#1{#1}\fi
\ifx\bissue    \undefined \def\bissue#1{#1}\fi
\ifx\bfpage    \undefined \def\bfpage#1{#1}\fi
\ifx\blpage    \undefined \def\blpage #1{#1}\fi
\ifx\burl      \undefined \def\burl#1{\textsf{#1}}\fi
\ifx\href      \undefined \def\href#1#2{\textsf{#2}}\fi
\ifx\betal     \undefined \def\betal{\textit{et al.}}\fi
\ifx\bctitle   \undefined \def\bctitle#1{#1}\fi
\ifx\beditor   \undefined \def\beditor#1{#1}\fi
\ifx\bbtitle   \undefined \def\bbtitle#1{\textit{#1}}\fi
\ifx\bedition  \undefined \def\bedition#1{#1}\fi
\ifx\bseriesno \undefined \def\bseriesno#1{\textbf{#1}}\fi
\ifx\blocation \undefined \def\blocation#1{#1}\fi
\ifx\bsertitle \undefined \def\bsertitle#1{\textit{#1}}\fi
\ifx\bsnm      \undefined \def\bsnm#1{#1}\fi
\ifx\bsuffix   \undefined \def\bsuffix#1{#1}\fi
\ifx\bparticle \undefined \def\bparticle#1{#1}\fi
\ifx\barticle  \undefined \def\barticle#1{}\fi
\ifx\binstitute  \undefined \def\binstitute#1{#1}\fi
\ifx\bpublisher  \undefined \def\bpublisher#1{#1}\fi
\ifx\doiurl    \undefined
  \def\doiurl#1{\href{http://dx.doi.org/#1}{\textsf{DOI}}}\fi
\ifx\arxivurl  \undefined
  \def\arxivurl#1{\href{http://arxiv.org/abs/#1}{\textsf{arXiv}}}\fi
\ifx\adsurl    \undefined
  \def\adsurl#1{\href{http://adsabs.harvard.edu/abs/#1}{\textsf{ADS}}}\fi
\ifx\botherref \undefined \def\botherref#1{}\fi
\ifx\url       \undefined \def\url#1{\textsf{#1}}\fi
\ifx\bchapter  \undefined \def\bchapter#1{}\fi
\ifx\bbook     \undefined \def\bbook#1{}\fi
\ifx\bcomment  \undefined \def\bcomment#1{#1}\fi
\ifx\oauthor   \undefined \def\oauthor#1{#1}\fi
\ifx\citeauthoryear \undefined\def \citeauthoryear#1{#1}\fi
\def\endbibitem {}
\ifx\bconflocation  \undefined \def\bconflocation#1{#1} \fi

\bibitem[\protect\citeauthoryear{{Brueckner}
  \textit{et~al.}}{1995}]{Brueckner1995}
\begin{barticle}
\bauthor{\bsnm{{Brueckner}}, \binits{G.E.}},
\bauthor{\bsnm{{Howard}}, \binits{R.A.}},
\bauthor{\bsnm{{Koomen}}, \binits{M.J.}},
\bauthor{\bsnm{{Korendyke}}, \binits{C.M.}},
\bauthor{\bsnm{{Michels}}, \binits{D.J.}},
\bauthor{\bsnm{{Moses}}, \binits{J.D.}},
\bauthor{\bsnm{{Socker}}, \binits{D.G.}},
\bauthor{\bsnm{{Dere}}, \binits{K.P.}},
\bauthor{\bsnm{{Lamy}}, \binits{P.L.}},
\bauthor{\bsnm{{Llebaria}}, \binits{A.}},
\bauthor{\bsnm{{Bout}}, \binits{M.V.}},
\bauthor{\bsnm{{Schwenn}}, \binits{R.}},
\bauthor{\bsnm{{Simnett}}, \binits{G.M.}},
\bauthor{\bsnm{{Bedford}}, \binits{D.K.}},
\bauthor{\bsnm{{Eyles}}, \binits{C.J.}}:
\byear{1995},
\batitle{{The Large Angle Spectroscopic Coronagraph (LASCO)}}.
\bjtitle{\solphys}
\bvolume{162},
\bfpage{357}.
\doiurl{10.1007/BF00733434}.
\end{barticle}
\endbibitem

\bibitem[\protect\citeauthoryear{{D{\'e}moulin} and
  {Aulanier}}{2010}]{Demoulin2010}
\begin{barticle}
\bauthor{\bsnm{{D{\'e}moulin}}, \binits{P.}},
\bauthor{\bsnm{{Aulanier}}, \binits{G.}}:
\byear{2010},
\batitle{{Criteria for flux rope eruption: Non-equilibrium versus torus
  instability}}.
\bjtitle{\apj}
\bvolume{718},
\bfpage{1388}.
\doiurl{10.1088/0004-637X/718/2/1388}.
\adsurl{2010ApJ...718.1388D}.
\end{barticle}
\endbibitem

\bibitem[\protect\citeauthoryear{{Efron}}{1982}]{Efron1982}
\begin{bbook}
\bauthor{\bsnm{{Efron}}, \binits{B.}}:
\byear{1982},
\bbtitle{{The Jackknife, the Bootstrap and Other Resampling Plans}}.
\bpublisher{CBMS-NSF Regional Conference Series in Applied Mathematics,
  {\bf 38}, Society for Industrial and Applied Mathematics (SIAM)},
\blocation{Philadelphia}.
\adsurl{1982jbor.book.....E}.
\end{bbook}
\endbibitem

\bibitem[\protect\citeauthoryear{{Forbes} and {Priest}}{1984}]{Forbes1984}
\begin{bchapter}
\bauthor{\bsnm{{Forbes}}, \binits{T.G.}},
\bauthor{\bsnm{{Priest}}, \binits{E.R.}}:
\byear{1984},
\bctitle{{Reconnection in solar flares}}.
In: \beditor{\bsnm{{Butler}}, \binits{D.M.}},
\beditor{\bsnm{{Papadopoulos}}, \binits{K.}} (eds.)
\bbtitle{Solar Terrestrial  Physics: Present and Future,
  NASA Reference Publ. {\bf 1120}.},
\bfpage{1}.
\end{bchapter}
\endbibitem

\bibitem[\protect\citeauthoryear{{Gopalswamy}
  \textit{et~al.}}{2010}]{Gopalswamy2010}
\begin{bchapter}
\bauthor{\bsnm{{Gopalswamy}}, \binits{N.}},
\bauthor{\bsnm{{Akiyama}}, \binits{S.}},
\bauthor{\bsnm{{Yashiro}}, \binits{S.}},
\bauthor{\bsnm{{M{\"a}kel{\"a}}}, \binits{P.}}:
\byear{2010},
\bctitle{{Coronal mass ejections from sunspot and non-sunspot regions}}.
In: \beditor{\bsnm{{Hasan}}, \binits{S.S.}},
\beditor{\bsnm{{Rutten}}, \binits{R.J.}} (eds.)
\bbtitle{Magnetic Coupling between the Interior and Atmosphere of the Sun},
\bpublisher{Springer},
\blocation{Berlin, Heidelberg},
\bfpage{289}.
\doiurl{10.1007/978-3-642-02859-5_24}.
\adsurl{2010mcia.conf..289G}.
\end{bchapter}
\endbibitem

\bibitem[\protect\citeauthoryear{Gosling}{1993}]{Gosling1993}
\begin{barticle}
\bauthor{\bsnm{Gosling}, \binits{J.T.}}:
\byear{1993},
\batitle{The solar flare myth}.
\bjtitle{J.~Geophys.~Res.}
\bvolume{98}(\bissue{A11}),
\bfpage{18,937}.
\end{barticle}
\endbibitem

\bibitem[\protect\citeauthoryear{{Handy} \textit{et~al.}}{1999}]{Handy1999}
\begin{barticle}
\bauthor{\bsnm{{Handy}}, \binits{B.N.}},
\bauthor{\bsnm{{Acton}}, \binits{L.W.}},
\bauthor{\bsnm{{Kankelborg}}, \binits{C.C.}},
\bauthor{\bsnm{{Wolfson}}, \binits{C.J.}},
\bauthor{\bsnm{{Akin}}, \binits{D.J.}},
\bauthor{\bsnm{{Bruner}}, \binits{M.E.}},
{\it et al.}:
\byear{1999},
\batitle{{The Transition Region and Coronal Explorer}}.
\bjtitle{\solphys}
\bvolume{187},
\bfpage{229}.
\doiurl{10.1023/A:1005166902804}.
\adsurl{1999SoPh..187..229H}.
\end{barticle}
\endbibitem

\bibitem[\protect\citeauthoryear{{Kliem} and {T{\"o}r{\"o}k}}{2006}]{Kliem2006}
\begin{barticle}
\bauthor{\bsnm{{Kliem}}, \binits{B.}},
\bauthor{\bsnm{{T{\"o}r{\"o}k}}, \binits{T.}}:
\byear{2006},
\batitle{{Torus instability}}.
\bjtitle{Phys. Rev. Lett.}
\bvolume{96}(\bissue{25}),
\bfpage{255002}.
\doiurl{10.1103/PhysRevLett.96.255002}.
\end{barticle}
\endbibitem

\bibitem[\protect\citeauthoryear{{Liu}}{2008}]{Liu2008}
\begin{barticle}
\bauthor{\bsnm{{Liu}}, \binits{Y.}}:
\byear{2008},
\batitle{{Magnetic field overlying solar eruption regions and kink and torus
  instabilities}}.
\bjtitle{\apjl}
\bvolume{679},
\bfpage{L151}.
\doiurl{10.1086/589282}.
\end{barticle}
\endbibitem

\bibitem[\protect\citeauthoryear{{Ma} \textit{et~al.}}{2010}]{Ma2010}
\begin{barticle}
\bauthor{\bsnm{{Ma}}, \binits{S.}},
\bauthor{\bsnm{{Attrill}}, \binits{G.D.R.}},
\bauthor{\bsnm{{Golub}}, \binits{L.}},
\bauthor{\bsnm{{Lin}}, \binits{J.}}:
\byear{2010},
\batitle{{Statistical study of coronal mass ejections with and without distinct
  low coronal signatures}}.
\bjtitle{\apj}
\bvolume{722},
\bfpage{289}.
\doiurl{10.1088/0004-637X/722/1/289}.
\adsurl{2010ApJ...722..289M}.
\end{barticle}
\endbibitem

\bibitem[\protect\citeauthoryear{{Mays} \textit{et~al.}}{2015}]{Mays2015}
\begin{barticle}
\bauthor{\bsnm{{Mays}}, \binits{M.L.}},
\bauthor{\bsnm{{Taktakishvili}}, \binits{A.}},
\bauthor{\bsnm{{Pulkkinen}}, \binits{A.}},
\bauthor{\bsnm{{MacNeice}}, \binits{P.J.}},
\bauthor{\bsnm{{Rast{\"a}tter}}, \binits{L.}},
\bauthor{\bsnm{{Odstrcil}}, \binits{D.}},
{\it et al.}:        
\byear{2015},
\batitle{{Ensemble Modeling of CMEs Using the WSA-ENLIL+Cone Model}}.
\bjtitle{\solphys}
\bvolume{290},
\bfpage{1775}.
\doiurl{10.1007/s11207-015-0692-1}.
\adsurl{2015SoPh..290.1775M}.
\end{barticle}
\endbibitem

\bibitem[\protect\citeauthoryear{{Poletto} and {Kopp}}{1986}]{Poletto1986}
\begin{bchapter}
\bauthor{\bsnm{{Poletto}}, \binits{G.}},
\bauthor{\bsnm{{Kopp}}, \binits{R.A.}}:
\byear{1986},
\bctitle{{Macroscopic electric fields during two-ribbon flares}}.
In: \beditor{\bsnm{{Neidig}}, \binits{D.F.}} (ed.)
\bbtitle{The Lower Atmosphere of Solar Flares; Proceedings of the Solar Maximum
  Mission Symposium},
\bpublisher{National Solar Observatory},
\blocation{Sunspot, NM},
\bfpage{453}.
\adsurl{1986lasf.conf..453P}.
\end{bchapter}
\endbibitem

\bibitem[\protect\citeauthoryear{{Qiu} and {Yurchyshyn}}{2005}]{Qiu2005}
\begin{barticle}
\bauthor{\bsnm{{Qiu}}, \binits{J.}},
\bauthor{\bsnm{{Yurchyshyn}}, \binits{V.B.}}:
\byear{2005},
\batitle{{Magnetic reconnection flux and coronal mass ejection velocity}}.
\bjtitle{\apjl}
\bvolume{634},
\bfpage{L121}.
\doiurl{10.1086/498716}.
\adsurl{2005ApJ...634L.121Q}.
\end{barticle}
\endbibitem

\bibitem[\protect\citeauthoryear{{Qiu} \textit{et~al.}}{2004}]{Qiu2004}
\begin{barticle}
\bauthor{\bsnm{{Qiu}}, \binits{J.}},
\bauthor{\bsnm{{Wang}}, \binits{H.}},
\bauthor{\bsnm{{Cheng}}, \binits{C.Z.}},
\bauthor{\bsnm{{Gary}}, \binits{D.E.}}:
\byear{2004},
\batitle{{Magnetic reconnection and mass acceleration in flare-coronal mass
  ejection events}}.
\bjtitle{\apj}
\bvolume{604},
\bfpage{900}.
\doiurl{10.1086/382122}.
\end{barticle}
\endbibitem

\bibitem[\protect\citeauthoryear{{Qiu} \textit{et~al.}}{2007}]{Qiu2007}
\begin{barticle}
\bauthor{\bsnm{{Qiu}}, \binits{J.}},
\bauthor{\bsnm{{Hu}}, \binits{Q.}},
\bauthor{\bsnm{{Howard}}, \binits{T.A.}},
\bauthor{\bsnm{{Yurchyshyn}}, \binits{V.B.}}:
\byear{2007},
\batitle{{On the magnetic flux budget in low-corona magnetic reconnection and
  interplanetary coronal mass ejections}}.
\bjtitle{\apj}
\bvolume{659},
\bfpage{758}.
\doiurl{10.1086/512060}.
\adsurl{2007ApJ...659..758Q}.
\end{barticle}
\endbibitem

\bibitem[\protect\citeauthoryear{{Richardson} and
  {Cane}}{2011}]{Richardson2011}
\begin{barticle}
\bauthor{\bsnm{{Richardson}}, \binits{I.G.}},
\bauthor{\bsnm{{Cane}}, \binits{H.V.}}:
\byear{2011},
\batitle{{Geoeffectiveness (Dst and Kp) of interplanetary coronal mass
  ejections during 1995-2009 and implications for storm forecasting}}.
\bjtitle{Space Weather}
\bvolume{9},
\bfpage{7005}.
\doiurl{10.1029/2011SW000670}.
\end{barticle}
\endbibitem

\bibitem[\protect\citeauthoryear{{Robbrecht}, {Patsourakos}, and
  {Vourlidas}}{2009}]{Robbrecht2009}
\begin{barticle}
\bauthor{\bsnm{{Robbrecht}}, \binits{E.}},
\bauthor{\bsnm{{Patsourakos}}, \binits{S.}},
\bauthor{\bsnm{{Vourlidas}}, \binits{A.}}:
\byear{2009},
\batitle{{No trace left behind: STEREO observation of a coronal mass ejection
  without low coronal signatures}}.
\bjtitle{\apj}
\bvolume{701},
\bfpage{283}.
\doiurl{10.1088/0004-637X/701/1/283}.
\adsurl{2009ApJ...701..283R}.
\end{barticle}
\endbibitem

\bibitem[\protect\citeauthoryear{{Sakurai}}{1982}]{Sakurai1982}
\begin{barticle}
\bauthor{\bsnm{{Sakurai}}, \binits{T.}}:
\byear{1982},
\batitle{{Green's function methods for potential magnetic fields}}.
\bjtitle{\solphys}
\bvolume{76},
\bfpage{301}.
\end{barticle}
\endbibitem

\bibitem[\protect\citeauthoryear{Scherrer \textit{et~al.}}{1995}]{Scherrer1995}
\begin{barticle}
\bauthor{\bsnm{Scherrer}, \binits{P.H.}},
\bauthor{\bsnm{Bogart}, \binits{R.S.}},
\bauthor{\bsnm{Bush}, \binits{R.I.}},
\bauthor{\bsnm{Hoeksema}, \binits{J.T.}},
\bauthor{\bsnm{Kosovichev}, \binits{A.G.}},
\bauthor{\bsnm{Schou}, \binits{J.}},
{\it et al.}:
\byear{1995},
\batitle{The Solar Oscillations Investigation - Michelson Doppler Imager}.
\bjtitle{Sol.~Phys.}
\bvolume{162},
\bfpage{129 }.
\end{barticle}
\endbibitem

\bibitem[\protect\citeauthoryear{{Sterling} and {Moore}}{2001}]{Sterling2001c}
\begin{barticle}
\bauthor{\bsnm{{Sterling}}, \binits{A.C.}},
\bauthor{\bsnm{{Moore}}, \binits{R.L.}}:
\byear{2001},
\batitle{{Internal and external reconnection in a series of homologous solar
  flares}}.
\bjtitle{\jgr}
\bvolume{106},
\bfpage{25227}.
\doiurl{10.1029/2000JA004001}.
\adsurl{2001JGR...10625227S}.
\end{barticle}
\endbibitem

\bibitem[\protect\citeauthoryear{{Suzuki}, {Welsch}, and
  {Li}}{2012}]{Suzuki2012}
\begin{barticle}
\bauthor{\bsnm{{Suzuki}}, \binits{J.}},
\bauthor{\bsnm{{Welsch}}, \binits{B.T.}},
\bauthor{\bsnm{{Li}}, \binits{Y.}}:
\byear{2012},
\batitle{{Are decaying magnetic fields above active regions related to coronal
  mass ejection onset?}}
\bjtitle{\apj}
\bvolume{758},
\bfpage{22}.
\doiurl{10.1088/0004-637X/758/1/22}.
\adsurl{2012ApJ...758...22S}.
\end{barticle}
\endbibitem

\bibitem[\protect\citeauthoryear{{Svestka} and {Cliver}}{1992}]{Svestka1992}
\begin{bchapter}
\bauthor{\bsnm{{Svestka}}, \binits{Z.}},
\bauthor{\bsnm{{Cliver}}, \binits{E.W.}}:
\byear{1992},
\bctitle{{History and basic characteristics of eruptive flares}}.
In: \beditor{\bsnm{{Svestka}}, \binits{Z.}},
\beditor{\bsnm{{Jackson}}, \binits{B.V.}},
\beditor{\bsnm{{Machado}}, \binits{M.E.}} (eds.)
\bbtitle{Eruptive Solar Flares},
\bsertitle{IAU Colloq.},
\bseriesno{399},
\bfpage{1}.
\doiurl{10.1007/3-540-55246-4_70}.
\adsurl{1992LNP...399....1S}.
\end{bchapter}
\endbibitem

\bibitem[\protect\citeauthoryear{{Wang} and {Zhang}}{2007}]{Wang2007b}
\begin{barticle}
\bauthor{\bsnm{{Wang}}, \binits{Y.}},
\bauthor{\bsnm{{Zhang}}, \binits{J.}}:
\byear{2007},
\batitle{{A comparative study between eruptive X-class flares associated with
  coronal mass ejections and confined X-class flares}}.
\bjtitle{\apj}
\bvolume{665},
\bfpage{1428}.
\doiurl{10.1086/519765}.
\adsurl{2007ApJ...665.1428W}.
\end{barticle}
\endbibitem

\bibitem[\protect\citeauthoryear{{Webb} and {Howard}}{2012}]{Webb2012}
\begin{barticle}
\bauthor{\bsnm{{Webb}}, \binits{D.F.}},
\bauthor{\bsnm{{Howard}}, \binits{T.A.}}:
\byear{2012},
\batitle{{Coronal Mass Ejections: Observations}}.
\bjtitle{Living Rev. in Solar Phys.}
\bvolume{9},
\bfpage{3}.
\doiurl{10.12942/lrsp-2012-3}.
\adsurl{2012LRSP....9....3W}.
\end{barticle}
\endbibitem

\bibitem[\protect\citeauthoryear{{Welsch} and {Li}}{2008}]{Welsch2008b}
\begin{bchapter}
\bauthor{\bsnm{{Welsch}}, \binits{B.T.}},
\bauthor{\bsnm{{Li}}, \binits{Y.}}:
\byear{2008},
\bctitle{{On the origin of strong-field polarity inversion lines}}.
In: \beditor{\bsnm{{Howe}}, \binits{R.}},
\beditor{\bsnm{{Komm}}, \binits{R.W.}},
\beditor{\bsnm{{Balasubramaniam}}, \binits{K.S.}},
\beditor{\bsnm{{Petrie}}, \binits{G.J.D.}} (eds.)
\bbtitle{Subsurface and Atmospheric Influences on Solar Activity},
\bsertitle{ASP Conf. Ser.}
\bseriesno{383},
\bfpage{429}.
\adsurl{2008ASPC..383..429W}.
\end{bchapter}
\endbibitem

\bibitem[\protect\citeauthoryear{{Xu} \textit{et~al.}}{2012}]{Xu2012}
\begin{barticle}
\bauthor{\bsnm{{Xu}}, \binits{Y.}},
\bauthor{\bsnm{{Liu}}, \binits{C.}},
\bauthor{\bsnm{{Jing}}, \binits{J.}},
\bauthor{\bsnm{{Wang}}, \binits{H.}}:
\byear{2012},
\batitle{{On the relationship between the coronal magnetic decay index and
  coronal mass ejection speed}}.
\bjtitle{\apj}
\bvolume{761},
\bfpage{52}.
\doiurl{10.1088/0004-637X/761/1/52}.
\adsurl{2012ApJ...761...52X}.
\end{barticle}
\endbibitem

\bibitem[\protect\citeauthoryear{{Zhao} and {Dryer}}{2014}]{Zhao2014}
\begin{barticle}
\bauthor{\bsnm{{Zhao}}, \binits{X.}},
\bauthor{\bsnm{{Dryer}}, \binits{M.}}:
\byear{2014},
\batitle{{Current status of CME/shock arrival time prediction}}.
\bjtitle{Space Weather}
\bvolume{12},
\bfpage{448}.
\doiurl{10.1002/2014SW001060}.
\adsurl{2014SpWea..12..448Z}.
\end{barticle}
\endbibitem

\end{thebibliography}


\end{document}